\documentclass{aastex62}
\usepackage{graphicx, epstopdf}
\usepackage{float}
\usepackage{verbatim}
\restylefloat*{figure}

\begin{document}

\title{A Lack of Evidence for Global Ram-pressure Induced Star Formation in the Merging Cluster Abell 3266 }

\correspondingauthor{Mark J. Henriksen}
\email{henrikse@umbc.edu}

\author{Mark J. Henriksen}
\affiliation{University of Maryland, Baltimore County \\
Physics Department \\
1000 Hilltop Circle \\
Baltimore, MD USA}

\author{Scott Dusek}
\affiliation{University of Maryland, Baltimore County \\
Physics Department \\
1000 Hilltop Circle \\
Baltimore, MD USA}

\begin{abstract}

Interaction between the intracluster medium and the interstellar media of galaxies via ram-pressure stripping (RPS)
has ample support from both observations and simulations of galaxies in clusters. 
Some, but not all of the observations and simulations show a phase of increased
star formation compared to normal spirals. Examples of galaxies undergoing RPS induced star formation in clusters experiencing
a merger have been identified in high resolution optical images supporting the existence of a star formation
phase. We have selected Abell 3266 to search
for ram-pressure induced star formation as a $global$ property of a merging cluster. Abell 3266 (z = 0.0594)
is a high mass cluster that features a high velocity dispersion, an infalling subcluster near to the line of sight, and a strong shock front. These phenomena
should all contribute to making Abell 3266 an optimum cluster to see the global effects of RPS induced star formation.
Using archival X-ray observations and published optical data, we 
cross-correlate optical spectral properties ([OII, H$\beta$]), indicative of starburst and post-starburst, respectively with ram-pressure, $\rho$v$^{2}$, calculated from the X-ray and optical data. 
We find that post-starburst galaxies, classified as E+A, occur at a higher frequency in this merging cluster than in the Coma cluster and at a comparable rate to intermediate redshift clusters. This is consistent with increased star formation due to the merger. However, both starburst and post-starburst galaxies are equally likely to be in a low or high ram pressure environment. From this result we infer that the duration of the starburst phase must be very brief so that: (1) 
at any time only a small fraction of the galaxies in a high ram pressure environment show this effect, and (2) most post-starburst galaxies are in an environment of low ram pressure due too their continued orbital motion in the cluster.
\end{abstract}


\section{Introduction} 

An estimated 20\% of all galaxies are in clusters of galaxies \citep{bah88}.  These large structures are typically $\sim$3 million light years or more in radius and are roughly circular in projection.  A hot ($~$10$^{7-8}$K), X-ray emitting intracluster medium (ICM) permeates all galaxy clusters. The X-ray spectra from the plasma has a thermal
bremsstrahlung continuum with emission lines from high ionization states of elements such as O, Fe, Si, and S. These elements are formed in stars and their presence in the ICM implies a connection to star formation, which takes place predominately in galaxies \citep{jim19}. The elements have an average global abundance of 0.3 Solar for rich clusters with kT $\sim$5 keV \citep{dep07}.  However, cluster radial metallicity profiles show a central peak of Fe as well as other elements \citep{sim09}.  While studies of sedimentation of metals to the center of the cluster could account, at least in part, for the excess metals \citep{med14}, inclusion of the intracluster magnetic field gives a prohibitively long time for this
process \citep{sar88}. With the expected slow settling time via diffusive models and inhibition from a likely non-radial magnetic field, sedimentation would be inefficient. Therefore, much of the metal enhancement must have originated from local galaxy evolutionary processes that are more efficient in the cluster center. \citep{gun72} originally suggested that as galaxies move through the ICM, there is a possibility that their interstellar medium (ISM) could be removed through RPS. The high velocity of the galaxy through the ICM and the local density of the ICM, which increases toward the center, results in a ram pressure that is higher in the inner 500 kpc than the outer parts of the cluster \citep{smi10}. The increased metals in a region of generally higher gas density (and therefore higher ram-pressure) is consistent with RPS being associated with the metal enrichment. Though there are competing galaxy evolutionary processes that favor the cluster center \citep{hen96}, images of gas removal in galaxies by RPS are evident in the jelly-fish galaxies.

Modeling ram-pressure stripping of the hot coronae of galaxies in clusters is consistent with the trend of smaller cluster galaxy coronae compared to the hot X-ray halos of field galaxies. This indicates that at least some gas removal occurs within cluster environments for recently accreted galaxies \citep{vij17}. In fact, evidence of ram-pressure stripping of M86 in the Virgo cluster was seen in the X-ray soon after it was predicted \citep{for79}. Later observations mapped the structure of its 150 kpc tidal tail in the X-ray with high resolution \citep{ran08} providing details of the stripping process.
More recently, an ample number of observations showing that RPS occurs in galaxy clusters \citep{cro06}; \citep{mer13}; \citep{ken14} have become available.  Simulations and observations generally agree that the outer disk ($>$ 20 kpc) is stripped first and the gas forms a tail in the wake of the galaxy \citep{gul17}; \citep{roe07}; \citep{ton10}; \citep{sun07}; \citep{fum14}; \citep{cor06}. Simulations are in general agreement that ram pressure increases the galactic star formation rate \citep{kro08}; \citep{kap08} and there are examples of increased star formation observed during RPS \citep{ram19}; \citep{owe12}; \citep{ebl14}; \citep{pog15}; \citep{koo04}. \citep{pog15} report a significant enhancement in star formation due to RPS in a large sample of clusters using optical spectroscopy. However, there are counter examples of RPS where enhanced star formation is not seen; for example, in the Virgo cluster \citep{cro08}. 

A general feature of simulations that are characterized by enhanced star formation is that ram pressure causes a compression of the gas within the disk of the galaxy. This transfer of gas from the outer regions of the galaxy to the inner, increases the gas density, resulting in an increase in the star formation rate. However, other simulations do not see enhanced star formation in the disk \citep{ton12} or only a slight enhancement for galaxies with select physical parameters and orientation \citep{tro16}; \citep{bek14}; \citep{ste16}.  Simulations of RPS during a galaxy cluster merger \citep{kro08}; \citep{kap08} show increased star formation due to local enhancements in gas density from the cluster wide shocks that typically accompany mergers and the broadened velocity dispersion. This phase of enhanced star formation in the disk and gas halo stripping \citep{bek03} lead to an overall reduction of the gas available to the galaxy, resulting in a subsequent decrease in the overall star formation rate. Thus, there is an optimum time to observe the increased star formation rate as it should only be enhanced for a relatively short time period after ram pressure stripping begins \citep{kro08}. The length of time depends on the strength of the ram pressure acting on the galaxy and the velocity of the galaxy, since this would affect the length of the RPS phase. As the galaxy encounters denser regions of the ICM, and hence stronger ram pressure, higher star formation rates should occur. This is directly observed in the Jellyfish galaxies found in merging clusters \citep{vul18}; \citep{vij13}; \citep{mcp16}; \citep{ebl14}. The increase in star formation should be accompanied by an observed enhancement in the equivalent width of emission or absorption lines.  The stripping timescale in the center of the cluster is estimated at 200 - 500 Myr \citep{ste16}; \citep{gul17}; \citep{smi10} so that if a galaxy cluster is observed after this time, any enhancement in these lines may no longer be observable.

\citep{pog15} has reported increased star formation via RPS in clusters of various morphology using a spectroscopic data set. In this paper, we will specifically address the connection with cluster merger. We have chosen a cluster with an ongoing merger to look for a global effect of increased star formation since there is some disagreement
about the importance of this phase in the simulations. Abell 3266 is a well studied cluster with an ongoing merger \citep{fin06}; \citep{hen02}; \citep{hen00}; \citep{deh17}. The cluster has a high velocity dispersion in the central region, 1367 km s$^{-1}$, while the outer region of the cluster exhibits a velocity dispersion similar to other rich clusters, $~$1000 km s$^{-1}$ \citep{hen02}. These velocities are consistent with the trend in values derived by \citep{deh17} though smaller
as these authors use a less restrictive velocity criterion for cluster membership and therefore model the larger cluster environment. The merger process creates more high velocity galaxies in the cluster as the kinetic energy of the infalling subcluster is turned into internal cluster energy via dynamical friction thus inflating the velocity dispersion in the central region. The cluster velocity distribution also has a high velocity component identified with a subcluster,  which contains $~$30 galaxies \citep{hen00}. Further evidence of a merger is an asymmetric gas distribution and regions of shocked gas \citep{hen02}. There should be higher ram pressure in these local regions of increased galaxy velocity and post-shock, higher density gas. Depending on the timescale of stripping, as discussed above, we expect to see a correlation
between high local ram pressure and the equivalent width of [OII] or H$\beta$ due to an increase in star formation. If no correlation is seen between ram pressure and these lines, it could be that this phase of star formation has already ceased. However, a cluster with an ongoing merger is the best case for catching the relatively short duration process of induced star formation.

\section{X-ray and Optical Data} 

Abell 3266 is a galaxy cluster located in the Southern Hemisphere at 4h31m24.1s, -61$\fd$26$\fm$38.0$fs$ (J2000) at a redshift of z=0.0594 \citep{deh17}. X-ray observations showed a central temperature enhancement in the intracluster medium, possibly due to a merger \citep{gra99}. A detailed temperature map confirmed the merger and provided details of the shock structure as well as the geometry of the merger \citep{hen02}. The extended cluster region shows a complex gas \citep{fin06} and optical structure containing six groups and
filaments to the north of the cluster \citep{deh17}. Observed with the ROSAT Position-Sensitive Proportional Counter (PSPC), Abell 3266 was observed for 13,547 seconds from August 19, 1993 to September 27, 1993. We choose to use the ROSAT PSPC over higher resolution data sets because its large
field of view, 2 degrees, matches the cluster size and provides full coverage of the Abell 3266 cluster.  The position, velocity, and equivalent width of H$\beta$ and [OII] for each galaxy was obtained from the Wide-Field Nearby Galaxy Clusters Survey \citep{cav09}. The spectroscopic data given in Table 1 is from \citep{fri14}. Typically, the equivalent width of H$\alpha$ is used as an optical indicator of recent star formation \citep{ken98}. When the spectroscopic data collected for Abell 3266 was fit, the H$\alpha$ line was found to be in a cluttered region of the spectrum. This is due to the presence of the NII spectral lines, at wavelengths of 6548{\AA} and 6583{\AA}, on either side of the H$\alpha$ line. The uncertainty in the H$\alpha$ line for this cluster lead us to choose the H$\beta$ line as the preferred indicator of star formation. When examining the intensity of the Balmer series transitions, the ratio of intensity of H$\alpha$ compared to H$\beta$ at 10,000 K is 2.87, and all subsequent Balmer transitions higher than H$\beta$ are substantially weaker. So, although H$\alpha$ could not be used for this study, H$\beta$ is a reasonable alternative for detecting star formation since it is the next strongest Balmer line. The [OII] emission is due to hot, young stars, namely O and B type stars. The optimal conditions for producing [OII] emission can be found in HII regions \citep{hog98}. Because of the short lifetime for these massive stars, [OII] emission is a good indicator of the most recent star formation.

There are other star formation indicators than those used here and indicators exist across the electromagnetic spectrum, including: radio 
line and continuum, FIR, optical emission and absorption lines, UV, and X-ray. Each indicator
has advantages and disadvantages  with regard to applicability and calibration (see references in \citep{ken98}).Indicators in the UV/optical/NIR 
rely on radiation from the stars themselves, while the FIR, radio, and X-ray are less direct. 
The radio line-to-continuum ratio, involving free-free emission from a star-forming region, must be separated from emission from contaminating 
dust and synchrotron emission (references in \citep{cal12}). The FIR, UV, and X-ray are associated with high mass stars and therefore recent or ongoing star formation, in the last 100 million years or so.
The Balmer and oxygen emission and absorption lines, used in this paper,  have the advantage of covering 
a broader range of stellar mass, from intermediate to high and therefore cover several billion years of star formation. However,
the optical lines have the disadvantage of sensitivity to attenuation by dust.

\section{Analysis}

Given the large number of groups in the cluster found optically that also appear to have enhanced X-ray
emission \citep{deh17}, we calculate the local gas density for each galaxy. The local gas density is obtained from an annulus with inner radius of 25 kpc and outer radius of 50 kpc. These values are nominally chosen to both exclude emission from a disk similar to the Milky Way and restrict gas density to the local value encountered by the galaxy. The angular scale for Abell 3266 is 70.4 kpc arcmin$^{-1}$ so the annulus dimensions are sub arc minute. The point-spread function for the PSPC is described by a Gaussian with a width that varies with distance from the detector center from 15 arc sec to 50 arc sec at 1 keV \citep{has92}. Thus the regions are folded through the point spread function (see Table 2). The energy range for each region was restricted to 0.5-2.4 keV, which maximizes source counts while minimizing background contamination. Only galaxies with an X-ray signal-to-noise count rate greater than 3 were used to obtain a gas density. The count rate is converted to a flux using an APEC model with specified values for temperature, abundance, and column density. The fluxes and normalizations are shown in Table 3. The ion density is calculated from the normalization, which contains the emission integral, using the sampled volume and an average electron per ion corresponding to the assumed abundances. The ion density is given in Table 4. The abundances assumed for the count rate to flux conversion as well as the density calculation is 90\% H and 10\% He, with metals ignored. The emission weighted PSPC temperature, 5.8 keV \citep{hen00} is used throughout the cluster as the PSPC is insensitive to temperature changes in the X-ray emitting gas. The column density is set to 2$\times$10$^{20}$ cm$^{-2}$ \citep{bek16}.

\begin{figure}[ht!]
\centering
\includegraphics[width=0.5\textwidth]{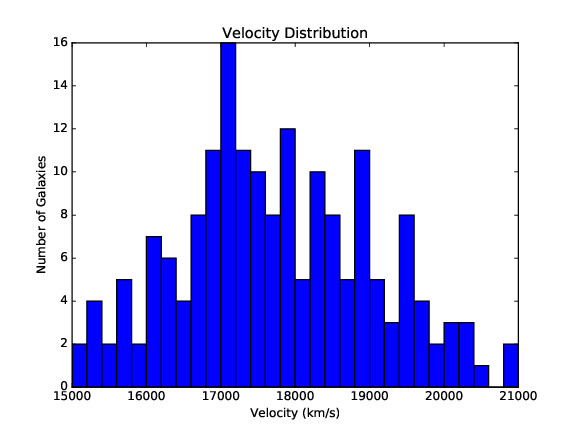}
\caption{Histogram of the galaxy velocities. The zero point is the velocity of the cluster, 17,804 km s$^{-1}$, z = 0.0594. The bin width is 200 km s$^{-1}$. Significant asymmetry is visible due
due the merger.}
\end{figure}

The ram pressure depends on the velocity of each galaxy in addition to
the ion density and mass. \citep{hen00} found that the range of galaxies in the Abell 3266, following a clipping procedure described in detail there, includes galaxies with heliocentric velocity between 15,000 and 21,000 km s$^{-1}$. Figure 1 shows the velocity histogram for the galaxies used in our analysis.

Of the original 263 galaxies in the WINGS survey, 51 are outside of the velocity range and are likely foreground or background. Another 35 galaxies are eliminated due to low signal-to-noise X-ray data or non-physical [OII] values. The heliocentric velocity of the main cluster core found by \citep{hen00} is 17,804 km s$^{-1}$ giving a redshift of 0.0594, consistent with \citep{deh17}. We use 0.0594 as the redshift throughout the paper. The velocity of each galaxy is found by subtracting the galaxy velocity from the heliocentric cluster velocity. The galaxy velocities are given in Table 4. The distribution of galaxies, as seen in Figure 1, is asymmetric and the peak is offset from zero, further evidence of the merger. 

In general, the gas density is higher toward the center, and lower in the outskirts. However, in a cluster merger, there is an asymmetry in the X-ray surface brightness, as can be seen in Figure 2. That is why a simple radial analysis, which assumes azimuthal symmetry is insufficient, justifying the analysis here that uses local density, to reflect gas distribution asymmetries from the merger. 

\section{Results} 

The ram pressure for each galaxy, $\rho$v$^{2}$, was calculated using the gas densities and velocities in Table 4. Figure 3 shows the histogram of ram pressure for the 177 galaxies in this study. As can be seen by observing the histogram, the galaxies experience a large range of ram pressure. We adopted the condition for high ram pressure used by \citep{kap09} of 5$\times$ 10$^{-11}$ dynes cm$^{-2}$. Of the 177 galaxies used, 61 galaxies were found to have ram pressures that can be considered “high” (i.e. $\geq$ 5 $\times$ 10$^{-11}$ dyn cm$^{-2}$. The remaining 116 galaxies were found to have “low” ram pressures. The calculated ram pressures for each of the 177 galaxies are listed in Table 4. We would expect to see galaxies with high ram pressures in regions where the density is high, or in galaxies with high velocities with respect to the cluster. In Figure 2, the distribution of these galaxies can be seen overlaid on the x-ray image of the cluster. Galaxies with low and high ram pressure appear throughout the cluster though high ram pressure galaxies tend to be closer to the center.

\begin{figure}[ht]
\begin{minipage}[b]{.4\textwidth}
\centering
\includegraphics[width=1\textwidth]{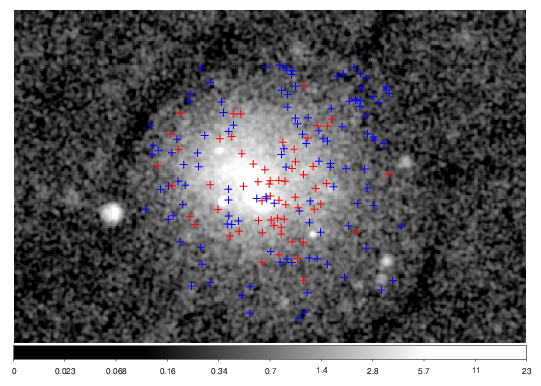}
\caption{Positions of galaxies overlaid on the x-ray image of Abell 3266 color coded to show ram pressure strength. Red depicts a galaxy with ram pressure $>$ 5 $\times$ 10$^{-11}$ dyn cm$^{-2}$ and blue a galaxy with ram pressure $\leq$ 5$\times$ 10$^{-11}$ dyn cm$^{-2}$.}
\end{minipage}
\hfill
\begin{minipage}[b]{.5\textwidth}
\centering
\includegraphics[width=1\textwidth]{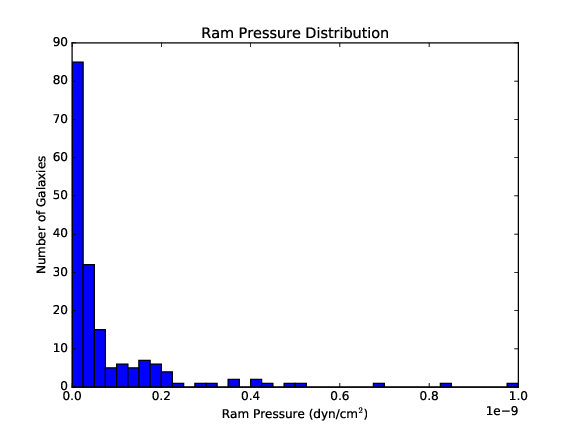}
\caption{Histogram of the ram pressure for each galaxy. The first two bins contain all galaxies with low ram pressure}
\end{minipage}
\end{figure}

Star formation in galaxies is enhanced by the ram pressure they experience. This is seen in both simulations and observations.
To assess whether a galaxy is currently experiencing or recently experienced an enhancement in star formation, we inspected the equivalent width of H$\beta$ and [OII] for each galaxy in the study. Figure 4 and 5 show the histogram of equivalent width of H$\beta$ and [OII],
respectively. 

\begin{figure}[ht]
\begin{minipage}[b]{.5\textwidth}
\centering
\includegraphics[width=1\textwidth]{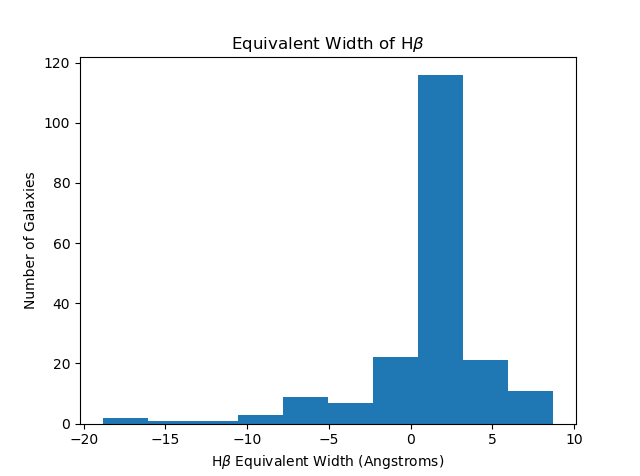}
\caption{Histogram of the equivalent width of H$\beta$. Each bin has a width of 0.5{\AA}.}
\end{minipage}
\hfill
\begin{minipage}[b]{.5\textwidth}
\centering
\includegraphics[width=1\textwidth]{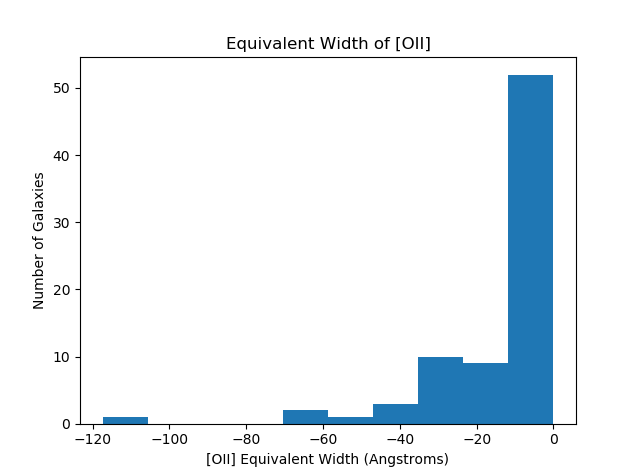}
\caption{Histogram of the equivalent width of [OII] emission lines. Each bin has a width of 2.5{\AA}.}
\end{minipage}
\end{figure}

It is believed that the surrounding environment can impact how the galaxy evolves (e.g. \citep{aba99}; \citep{mil03}; \citep{roe07}; \citep{fri11}; \citep{fri14}. 
If there is a correlation between ram pressure and star formation in Abell 3266, it should be apparent when comparing a histogram of equivalent widths for galaxies with high ram pressures verses galaxies with low ram pressures. Figures 6 and 7 show the histograms of the equivalent width of H$\beta$ and [OII], respectively, plotted based on whether the galaxy has high or low ram pressure. Note that If the equivalent width of H$\beta$ is positive (negative), this corresponds to absorption (emission). Perhaps there is a weak correlation that is buried in the large, inclusive samples. 

\begin{figure}[ht]
\begin{minipage}[b]{.5\textwidth}
\centering
\includegraphics[width=1\textwidth]{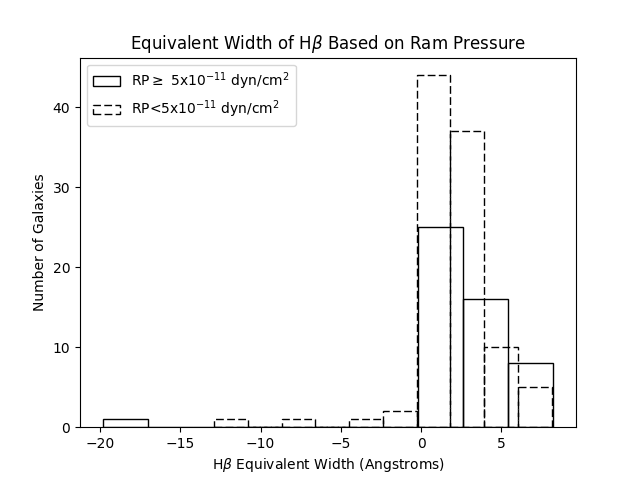}
\caption{Histogram of H$\beta$ equivalent width based on the ram pressure of the galaxy. The solid histogram corresponds to high ram pressure ($\geq$ 5 x 10$^{-11}$ dyn cm$^{-2}$). The black dashed histogram corresponds to low ram pressure ($<$ 5 x 10$^{11}$ dyn cm$^{-2}$). Each bin has a width of 0.5{\AA}. The low and high ram pressure galaxies span the same range of
absorption line strength though the high absorption line galaxies have a higher fraction
of high ram pressure galaxies. Low absorption line galaxies are preferentially low ram pressure.}
\end{minipage}
\hfill
\begin{minipage}[b]{.5\textwidth}
\centering
\includegraphics[width=1\textwidth]{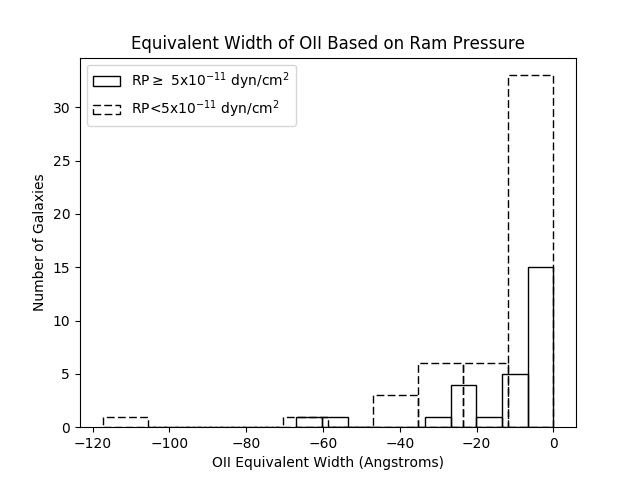}
\caption{Histogram of [OII] equivalent width based on the ram pressure of the galaxy. The solid histogram corresponds to high ram pressure($\geq$ 5 x 10$^{-11}$ dyn cm$^{-2}$). The black dashed histogram corresponds to low ram pressure ($<$ 5 x 10$^{11}$ dyn cm$^{-2}$). Each bin has a width of 2.5{\AA}.}
\end{minipage}
\end{figure}

\begin{figure}[ht]
\begin{minipage}[b]{.5\textwidth}
\centering
\includegraphics[width=1\textwidth]{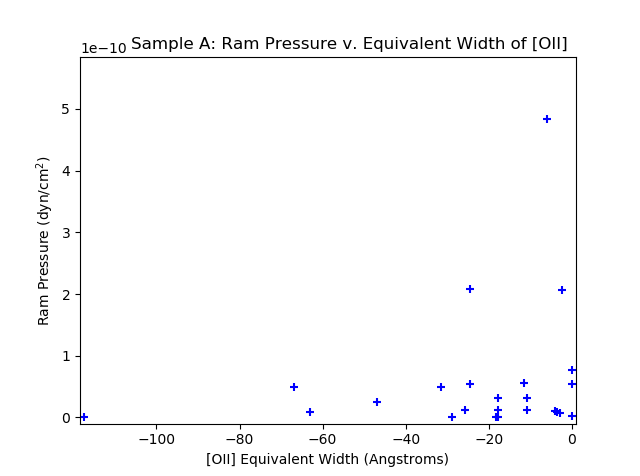}
\caption{Plot of equivalent width of [OII] verses ram pressure for Sample A.}
\end{minipage}
\hfill
\begin{minipage}[b]{.5\textwidth}
\centering
\includegraphics[width=1\textwidth]{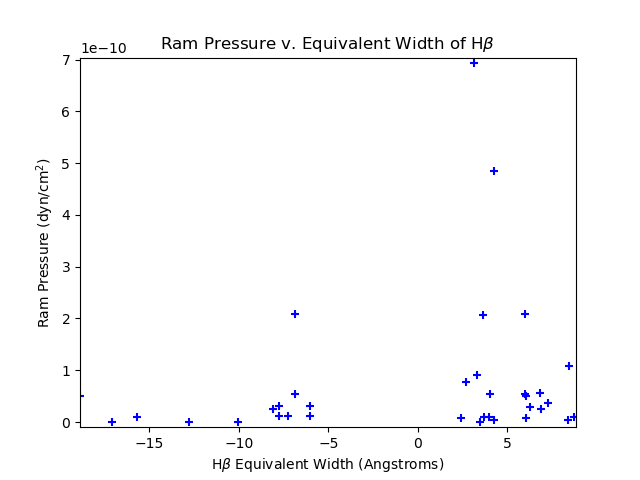}
\caption{Plot of equivalent width of H$\beta$ verses ram pressure for sample A, the 34 galaxies with 3 sigma significance. There is
no visible preference for galaxies with H$\beta$ in emission to be in a region of high ram pressure.} 
\end{minipage}
\end{figure}

To evaluate this hypothesis, we addressed the fact that
the equivalent widths of H$\beta$ and [OII] obtained from WINGS do not all have a signal to noise of 3$\sigma$ or greater. For this reason we have created two subsamples which will be referred to as Sample A and Sample B. Sample A consists of all galaxies with line emission greater than 3$\sigma$ significance. Sample A consists of a total of 34 galaxies for H$\beta$ and 24 for O[II]. Figures 8 and 9 show a plot of O[II] verses ram pressure and H$\beta$  verses ram pressure, respectively, for the Sample A galaxies. Sample B is a subgroup of Sample A, consisting of all galaxies with a signal to noise greater than 3$\sigma$ for both H$\beta$ and [OII].  Sample B consists of a total of 12 galaxies. Figures 10 and 11 show a plot of H$\beta$ verses ram pressure and [OII] verses ram pressure, respectively for Sample B galaxies. Sample B is the highest quality subsample. However, in all of the data samples, there is no apparent correlation between H$\beta$ or O[II] line strength and ram-pressure.

\begin{figure}[ht]
\begin{minipage}[b]{.5\textwidth}
\centering
\includegraphics[width=1\textwidth]{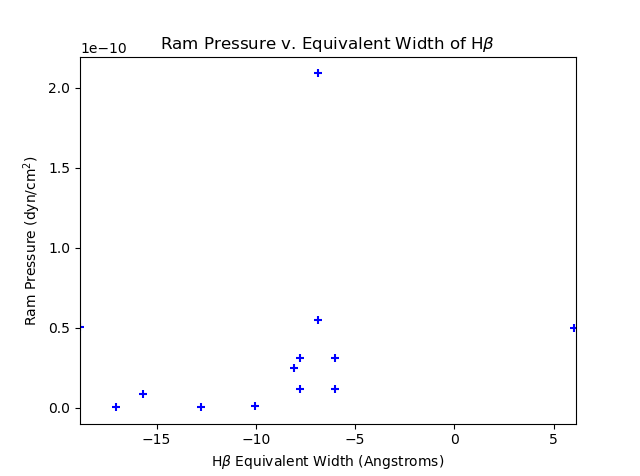}
\caption{Log-log plot of the equivalent width of H$\beta$ verses ram pressure for the 12 galaxies (Sample B) with both H$\beta$ and O[II] lines of higher than 3$\sigma$ significance.}
\end{minipage}
\hfill
\begin{minipage}[b]{.5\textwidth}
\centering
\includegraphics[width=1\textwidth]{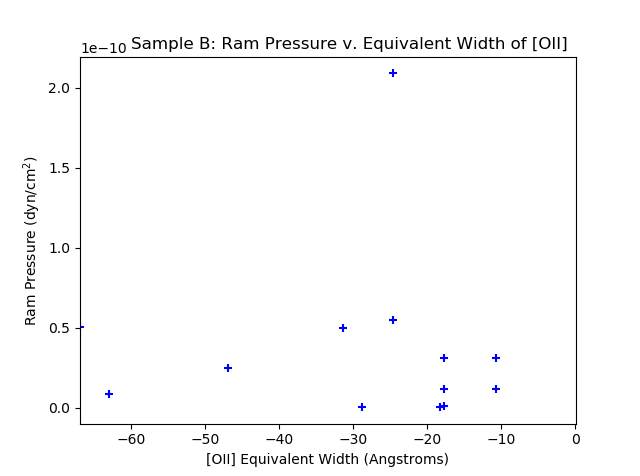}
\caption{Log-log plot of the equivalent width of O[II] verses ram pressure for 12 galaxies with both H$\beta$ and O[II] lines of higher than 3$\sigma$ significance.}
\end{minipage}
\end{figure}




\subsection{E+A Galaxies}

\citep{fis98} defines an E+A galaxy as one with Balmer absorptions lines that are (H$\delta$+H$\gamma$+H$\beta$)/3 $>$ 4{\AA}, and [OII] emission lines are $<$ 5{\AA}. For our criteria, we come as close as we can to this with our database by requiring an equivalent width of either H$\beta$ or H$\delta$ $>$ 4{\AA}, and equivalent width of [OII] $<$ -5{\AA}. There are 6 galaxies out of the 71 (that have spectroscopic coverage) or 8.5\% that meet these classification requirements. Of those 6, 3 have “high” ram pressures $>$ 5 x 10$^{-11}$, while the other 3 have “low” ram pressures. Therefore, the data do not support a correlation between the E+A phenomenon and ram pressure. However, the frequency of E+A galaxies, 8.5\% is substantially higher than the fraction found in a non-merging cluster, Coma \citep{cal93} and this indicates a galaxy evolutionary effect associated with the merger.

\section{Discussion}

The relationship between ram pressure and star formation during a cluster merger was investigated for Abell 3266. This galaxy cluster was chosen because it shows evidence of a merger in progress: very high velocity dispersion in the center, an asymmetric density distribution, and regions of shock heated gas. Thus, it is the best opportunity to catch an increase in star formation due to RPS before the galaxies are quenched. Spectral type O and B type stars emit continuum radiation that ionizes Hydrogen in the surrounding medium. The Hydrogen ions then recombine with free electrons producing recombination emission lines. These emission lines are believed to correspond to HII regions surrounding O and B type stars. Other stellar classes, such as A and F type stars, do not produce strong enough radiation to ionize the surrounding gas and thus are lacking Hydrogen emission lines. Because of the short lifetime of the high mass stars, this emission line should track starburst galaxies.
We found that the galaxies exhibiting H$\beta$ in emission do not correlate with high ram pressure in our study. We do not see evidence of a widespread RPS induced starburst phase associated with the merger. There may be evidence of an increased 
level of recent star formation due to the merger, but we don't find a direct link to ram pressure.
We then turn our attention to the possibility of post-starburst galaxies and their correlation with ram pressure.

An episode of rapid star formation would consume a large percentage of the gas available in the galaxy. This consumption, coupled with ram pressure stripping, would lead to a termination of star formation. While the brief timescale of enhanced star formation via RPS is critical to whether or not a correlation is observed it is possible to detect the post-starburst galaxies as E+A galaxies. An E+A galaxy has a characteristic spectrum, displaying strong Balmer absorption and little or no [OII] emission (\citep{tra03}; \citep{fri14}; \citep{cal93}. The star formation observed in an E+A galaxy is understood to have occurred within the last 1.5 Gyr \citep{cou87}. \citep{tra03} observed that the number of E+A galaxies found in intermediate redshift clusters is greater than the number found in the field. This implies that environment is important in their formation. Such events as ram pressure stripping could play a role in the increased number of post starburst galaxies found in clusters verses the field. \citep{tra03} and \citep{dre99} both observe that E+A’s are generally located outside of the cluster core, which may be a result of their continued motion after RPS induced starburst.

The [OII] emission comes from the medium surrounding hot, young stars, namely O and B type stars. The optimum conditions for producing [OII] emission can be found in HII regions \citep{hog98}.
This makes the [OII] emission line a good indication of young stellar populations. If the [OII] emission line is very weak, or absent, it is likely that there is no current star formation occurring in the galaxy. Pairing this information with strong Balmer absorption lines allows us to classify galaxies as post starburst. If there was a recent starburst in a galaxy, there will still be some O and B type stars present which have not reached the end of their lifetime ($~$1-10 Myr); however, there will be very little gas remaining in the former HII regions, resulting in a lack of [OII] emission. The A and F type stars, with lifetimes on the order of $~$1 Gyr or longer, will still provide strong Balmer absorption lines in the spectra of those galaxies. We find here that 8.5\% or 6 out of 71 of the galaxies that have both OII and H$\beta$ or H$\delta$ available are E+A galaxies.  \citep{tra03} find E+A's make up a non-negligible component of the cluster population (7\% - 13\%) at intermediate redshifts. These authors point out that compared to the low E+A fraction in Coma (3\%)\citep{cal93} their results show the E+A fraction evolves strongly with redshift. We find a comparable fraction, 8.5\% in Abell 3266, a low redshift merging cluster to the intermediate
redshift clusters. Within our limited
sample we can say the E+A fraction is higher in this merging cluster compared to the nearby Coma cluster. The 6 E+A galaxies are split evenly between high and low ram pressure. While this E+A analysis does not show a correlation between post starburst galaxies and high ram pressure, the lack of immediacy in this phenomena means that it is likely the E+A galaxies are observed in a different gas environment than the one they were formed in.


The lack of a strong global correlation between ram-pressure and optical star formation indicators may also be complicated by the multiplicity of cluster environmental effects on galaxy evolution. For example, radio observations reveal an asymmetric, truncated distribution of HI in the spiral galaxy, UGC 07049, in the NGC 4065 group of galaxies \citep{fre10}. In addition, radio continuum emission reveals strong star formation in the galaxy.
The authors conclude that a combination of tidal and ram-pressure stripping are most likely necessary to produce the HI deficiency and disk asymmetry. In a study of three Virgo Cluster galaxies, in one of the galaxies, NGC 4264, X-ray observations at the position of magnetic field enhancements do not show compressional heating as expected from a strong adiabatic shock due to ram pressure. This favors tidal interactions rather than ram pressure stripping. Conversely, NGC 4569’s radio polarized ridge shows a higher temperature, indicative of ram-pressure shocked gas. The third galaxy, NGC 2276, shows no clear indications of a shock, which is consistent with the observed distortions coming from tidal interaction \citep{wez12}.

Finally, though there is evidence of an ongoing merger in Abell 3266, it may have previously had a major merger at an earlier epoch that resulted in cluster wide quenching of star formation diminishing the effect of the present merger. Support for this idea comes from studies at 
intermediate redshift. \citep{vau20} find that star formation takes place throughout the stellar disk in redshift, 0.3 - 0.6 galaxies though cluster galaxies have a significantly smaller H$\alpha$ disk than field galaxies. The authors concluded that ram-pressure stripping and disk strangulation can account for this result. The high level of E+A galaxies we find is consistent with a larger merger within the last $\sim$billion years.

\section{Conclusions}

Through this work the properties of ram pressure and H$\beta$ equivalent width are examined for 177 galaxies located in the merging galaxy cluster Abell 3266. There are cases of ram pressure induced star formation within galaxies, the so called Jellyfish galaxies that appear to be associated with cluster merger. Although Abell 3266 is an ongoing merger the expected relationship between ram pressure and H$\beta$ equivalent width was not observed. This shows a lack of evidence for ongoing global star formation due to ram pressure at this point of the merger. To test whether star formation had recently ended, we identified galaxies whose spectra were characteristic of E+A galaxies, or post starburst galaxies. There did not appear to be a correlation with E+A galaxies and ram pressure either. However, there is an enhancement in the
number of E+A galaxies in this merging cluster compared to a quiescent cluster or the field, which is consistent with a major merger in the past billion years. The evidence contained in this work does not provide clear support that star formation is currently enhanced globally by ram pressure globally. But because simulations show that the enhanced star formation phase is short and depends sensitively on the environment conditions the galaxy encounters, we conclude that this phase is episodic, involving only
a small fraction of the cluster galaxies at any epoch. The lack of correlation is also likely the result of the galaxy's continued motion after the RPS
induced star formation phase.


\startlongtable
\begin{deluxetable}{ccccccccc}
	\tabletypesize{\scriptsize}
	\tablecolumns{9}
	\tablenum{1}
	\tablewidth{0pt}
	\tablehead{
		 \colhead{WINGS ID} & \colhead{[OII] EW} & \colhead{[OII] EW Error} & \colhead{H$\delta$ EW} & \colhead{H$\delta$ EW Error} & \colhead{H$\beta$ EW} &  \colhead{H$\beta$ EW Error} &  \colhead{RA} & \colhead{DEC}\\
		 \colhead{}  & \colhead{Width ($\AA$)} & \colhead{Width ($\AA$)} & \colhead{Width ($\AA$)} & \colhead{Width ($\AA$)} & \colhead{Width ($\AA$)} & \colhead{Width ($\AA$)} & \colhead{Degrees} & \colhead{Degrees} 
		 }
	\startdata
J042847.65-613052.9 & - & - & 2.16 & 0.81 & 0.87 & 0.58 & 67.19854 & -61.51469 \\
J042856.57-613753.8&-&-&7.98&2.27&0.29&0.34&67.23571&-61.63161 \\ 
J042902.02-611354.9&-&-&2.2&1.15&2.88&1.15&67.25842&-61.23192 \\ 
J042902.33-612417.2&-9.37&2.59&5.29&1.25&-2.02&0.86&67.25971&-61.40478 \\
J042905.27-611302.4&-10.72&2.79&3.77&1.16&-7.76&1.7&67.27196&-61.21733 \\ 
J042905.22-611321.7&-17.76&2.54&7&1.79&-6&1.42&67.27175&-61.22269 \\
J042907.20-612018.3&-&-&1.03&0.6&2.55&0.88&67.28&-61.33842 \\
J042908.67-613906.4&-9.96&6.86&-2.81&1.04&1.92&0.86&67.28612&-61.65178 \\
J042912.95-611508.4&-35.65&8.31&4.76&1.28&1.67&0.69&67.30396&-61.25233 \\ 
J042916.95-611953.1&-11.09&1.9&1.82&1.16&-4.08&1.96&67.32062&-61.33142 \\ 
J042917.68-612818.2&-&-&1.06&0.63&1.41&0.71&67.32367&-61.47172 \\  
J042918.62-611929.8&-&-&1.83&0.8&0.84&0.49&67.32758&-61.32494 \\ 
J042918.93-611419.3&-&-&3.08&1.11&1.29&0.68&67.32887&-61.23869 \\ 
J042924.75-611905.7&-&-&0.79&0.51&1.99&0.7&67.35312&-61.31825 \\ 
J042924.92-612603.4&-&-&0.92&0.56&1.3&0.6&67.35383&-61.43428 \\ 
J042925.36-613345.9&-&-&0.83&0.68&0.73&0.53&67.35567&-61.56275 \\
J042926.28-611155.9&0&0.56&4.16&1.25&0.82&0.59&67.3595&-61.19886 \\
J042927.18-611651.9&-2.93&2.1&3.48&1.11&2.4&0.76&67.36325&-61.28108 \\
J042927.89-612337.0&-&-&1.44&0.71&2.45&0.91&67.36621&-61.39361 \\ 
J042932.77-611452.5&-&-&0.44&0.38&1.46&0.72&67.38654&-61.24792 \\
J042933.01-611124.4&-2.18&1.31&2.47&0.92&-1.02&1.08&67.38754&-61.19011 \\
J042933.10-611536.3&-2.55&1.53&1.73&0.77&0.74&0.46&67.38792&-61.26008 \\ 
J042934.78-612803.7&-&-&1.05&0.6&1.59&0.73&67.39492&-61.46769 \\  
J042936.13-611450.2&-&-&-&-&2.48&1.13&67.40054&-61.24728 \\  
J042937.98-611445.5&-&-&1.2&0.66&2.41&0.92&67.40825&-61.24597 \\ 
J042939.59-611034.5&-30.48&5.48&-0.24&0.27&-1.58&0.78&67.41496&-61.17625 \\ 
J042944.51-611456.3&-17.76&3.3&2.31&0.82&-10.04&1.99&67.43546&-61.24897 \\
J042944.77-613054.9&-&-&2.68&1.05&2.83&1.05&67.43654&-61.51525 \\ 
J042946.91-613025.6&-31.43&8.41&-7.68&1.45&6.02&1.59&67.44546&-61.50711 \\
J042947.62-612337.0&-&-&-&-&7.11&3.14&67.44842&-61.39361 \\ 
J042948.73-613732.2&-&-&-&-&2.44&0.87&67.45304&-61.62561 \\ 
J042951.03-611127.3&-&-&1.38&0.69&1.55&0.72&67.46262&-61.19092 \\ 
J042951.96-611812.9&-4.13&1.43&-&-&0.36&0.42&67.4665&-61.30358 \\
J042957.45-611145.7&-&-&1.26&0.72&2.57&0.98&67.48937&-61.19603 \\
J043000.74-612554.3&-21.87&16.42&2.23&1.08&1.98&1.07&67.50308&-61.43175 \\ 
J043003.09-611714.0&-22.73&8.03&2.3&1.13&0.89&0.96&67.51287&-61.28722 \\
J043003.77-611555.4&-&-&0.27&0.38&0.14&0.33&67.51571&-61.26539 \\ 
J043004.38-612255.5&-&-&3.78&1.13&2.32&1.06&67.51825&-61.38208 \\ 
J043004.47-612325.7&0&0.7&2.56&0.95&1.52&0.73&67.51862&-61.39047 \\
J043006.99-613551.2&-&-&5.8&1.79&4.2&1.49&67.52912&-61.59756 \\ 
J043007.64-612003.6&0&0.49&1.08&0.67&2.25&0.92&67.53183&-61.33433 \\ 
J043008.03-611809.3&-30.83&7.46&0.02&0.14&-1.23&1.24&67.53346&-61.30258 \\ 
J043008.45-612529.8&-&-&-&-&3.78&1.94&67.53521&-61.42494 \\
J043012.30-611946.4&-&-&-&-&-6.82&2.85&67.55125&-61.32956 \\ 
J043014.66-612808.7&-&-&7.34&1.47&3.29&1&67.56108&-61.46908 \\ 
J043014.72-613145.4&-&-&-&-&1.75&0.87&67.56133&-61.52928 \\
J043016.68-612239.4&-&-&-&-&3.02&1.9&67.5695&-61.37761 \\ 
J043017.05-611844.1&-46.92&5.74&4.78&1.25&-8.07&1.71&67.57104&-61.31225 \\ 
J043018.11-613508.8&-3.17&1.14&2.2&0.87&-2.12&0.84&67.57546&-61.58578 \\ 
J043018.94-611806.9&-&-&5.39&1.48&0.41&0.53&67.57892&-61.30192 \\ 
J043020.03-612609.7&-&-&3.09&1.18&2.48&1.02&67.58346&-61.43603 \\
J043021.56-612848.7&0&0.25&1.44&0.68&1.24&0.66&67.58983&-61.48019 \\ 
J043022.67-612319.6&-12.71&5.45&4.67&1.38&1.3&0.68&67.59446&-61.38878 \\
J043025.99-612745.9&-&-&5.28&2.55&-4.79&1.8&67.60829&-61.46275 \\ 
J043026.05-613524.8&-35.58&18.78&-12.89&1.98&2.69&0.94&67.60854&-61.59022 \\
J043026.74-613031.9&-4.99&2.82&1.71&0.83&2.82&1.09&67.61142&-61.50886\\
J043026.85-613503.1&-&-&8.14&1.93&7.24&2.18&67.61187&-61.58419 \\
J043028.36-611704.0&-17.28&5.12&1.49&0.86&0.43&0.74&67.61817&-61.28444 \\
J043028.68-611939.8&-&-&-&-&0.49&0.46&67.6195&-61.32772 \\ 
J043028.84-613932.5&-&-&2.92&1.19&2.58&1&67.62017&-61.65903 \\ 
J043030.43-612023.1&-3.23&1.61&2.89&1.02&0.51&0.46&67.62679&-61.33975 \\ 
J043032.07-614200.6&-25.8&8.63&0.03&0.48&-7.23&1.69&67.63362&-61.70017 \\
J043032.53-611257.5&-&-&2.46&1&0.67&0.82&67.63554&-61.21597 \\ 
J043032.85-613755.5&-&-&1.42&0.73&1.69&0.92&67.63687&-61.63208 \\ 
J043033.23-613300.7&-&-&4.77&2.09&8.44&2.63&67.63846&-61.55019 \\ 
J043034.31-611912.6&-&-&2.3&1.17&1.86&1.05&67.64296&-61.32017 \\ 
J043037.57-614249.6&-&-&1.85&0.97&6.42&2.62&67.65654&-61.71378 \\
J043037.69-612902.7&-&-&1.47&0.75&1.28&0.74&67.65704&-61.48408 \\ 
J043038.92-612837.5&-&-&2.14&0.94&3.38&1.42&67.66217&-61.47708 \\
J043039.53-613507.8&-&-&2.2&0.9&1.39&0.66&67.66471&-61.5855 \\
J043039.66-611751.9&-3.47&2.26&2.74&1.11&8.73&1.88&67.66525&-61.29775 \\
J043041.79-612243.1&-3.15&1.78&1.37&0.6&1.76&0.75&67.67412&-61.37864 \\
J043041.87-612715.8&0&0.6&0.91&0.54&2.7&0.81&67.67446&-61.45439 \\
J043041.86-611303.0&-&-&1.95&0.84&2.8&1.02&67.67442&-61.2175 \\
J043041.88-611709.9&-&-&-&-&2.05&0.82&67.6745&-61.28608 \\ 
J043043.25-612106.9&-66.91&7.36&-&-&-18.85&2.49&67.68021&-61.35192 \\ 
J043044.89-611104.6&-&-&-&-&1.5&0.85&67.68704&-61.18461 \\ 
J043045.32-613300.4&-&-&4.22&1.41&1.5&0.83&67.68883&-61.55011 \\ 
J043045.38-612335.7&0&0.37&1.38&0.63&0.54&0.37&67.68908&-61.39325 \\
J043045.64-613538.7&-&-&0.69&0.63&2.09&1.32&67.69017&-61.59408 \\  
J043046.27-611132.2&-&-&-&-&1.25&0.61&67.69279&-61.19228 \\
J043050.16-611543.9&0&0.09&1.5&0.71&1.9&0.8&67.709&-61.26219 \\
J043050.35-611406.0&0&1.18&-&-&1.53&1.03&67.70979&-61.235 \\ 
J043051.24-611059.7&0&0.26&1.07&0.52&0.94&0.55&67.7135&-61.18325 \\ 
J043051.87-612324.4&-&-&3.13&1.16&6.89&1.64&67.71612&-61.39011 \\ 
J043052.13-611034.7&-&-&-&-&-0.73&0.83&67.71721&-61.17631 \\
J043052.25-612813.8&-&-&1.89&0.77&0.59&0.46&67.71771&-61.4705 \\
J043053.08-612514.2&-3.05&2.26&4.28&1.13&3.2&1.57&67.72117&-61.42061 \\
J043053.73-613005.7&-&-&1.65&0.77&2.3&0.83&67.72387&-61.50158 \\ 
J043054.21-612103.2&-&-&2.05&0.82&1.48&0.69&67.72587&-61.35089 \\
J043055.49-612014.3&-&-&2.27&1.13&2.01&0.98&67.73121&-61.33731 \\ 
J043056.84-613202.3&-&-&1.53&0.71&1.74&0.71&67.73683&-61.53397 \\
J043056.87-611333.5&-&-&3.31&1.22&3.79&1.55&67.73696&-61.22597 \\
J043056.90-612147.4&-&-&3.66&1.34&-2.01&1.67&67.73708&-61.36317 \\ 
J043057.00-613107.3&-11.63&6.7&-19.85&2.91&6.8&2.09&67.7375&-61.51869 \\ 
J043058.25-613536.2&-5.99&3.96&0.73&0.47&0.21&0.54&67.74271&-61.59339 \\ 
J043058.75-611021.0&-&-&-&-&-1.28&0.8&67.74479&-61.1725 \\ 
J043059.59-612507.2&-&-&-&-&1.04&0.79&67.74829&-61.41867 \\ 
J043059.55-612740.1&-9.39&8.53&6.53&1.61&-9.28&4.98&67.74812&-61.46114 \\
J043059.77-613438.5&0&0.17&1.19&0.71&1.76&0.89&67.74904&-61.57736 \\ 
J043100.91-612957.9&-57.74&12.48&-&-&-5.96&4.46&67.75379&-61.49942 \\
J043104.72-613053.4&-&-&5.42&1.52&-0.7&1.02&67.76967&-61.51483 \\
J043104.61-612802.6&-&-&1.21&0.57&1.71&0.67&67.76921&-61.46739 \\ 
J043107.70-613012.1&-&-&3.53&1.35&-0.46&0.78&67.78208&-61.50336 \\
J043108.34-613413.8&-&-&4.39&1.36&1.17&1.01&67.78475&-61.5705 \\ 
J043108.71-612707.0&-2.82&2.08&2.41&0.94&1.69&1.01&67.78629&-61.45194 \\
J043108.81-612510.1&-24.6&7.22&3.67&1.01&-6.88&1.45&67.78671&-61.41947 \\
J043109.75-612535.8&-&-&-&-&5.96&1.83&67.79062&-61.42661 \\
J043112.64-612728.9&0&0.29&1.12&0.68&1.75&0.76&67.80267&-61.45803 \\
J043113.20-611037.9&-&-&-&-&0.64&0.65&67.805&-61.17719 \\
J043114.59-612344.7&0&1.49&1.97&1.09&1.83&0.95&67.81079&-61.39575 \\
J043117.02-612724.2&-&-&7.41&1.76&1.08&1.15&67.82092&-61.45672 \\
J043117.25-613149.8&-2.3&1.69&6.24&1.29&3.62&1.09&67.82187&-61.5305 \\
J043117.66-612743.7&-&-&-&-&3.37&1.2&67.82358&-61.46214 \\
J043121.54-613016.3&-&-&-&-&-0.85&1.14&67.83975&-61.50453 \\
J043121.86-612517.7&-&-&0.93&0.93&0.56&0.74&67.84108&-61.42158 \\
J043123.30-612726.9&-&-&1.31&0.65&0.52&0.42&67.84708&-61.45747 \\
J043126.94-612301.3&-5.97&2.29&8.25&1.62&4.28&1.07&67.86225&-61.38369 \\
J043130.61-613845.0&-26.95&4.34&2.65&0.89&-1.86&1.07&67.87754&-61.64583 \\
J043131.08-614208.8&-&-&1.53&1.03&3.91&1.49&67.8795&-61.70244 \\
J043137.44-612552.9&-&-&-&-&3.12&0.99&67.906&-61.43136 \\
J043139.30-613951.6&-&-&-&-&-4.61&1.81&67.91375&-61.66433 \\
J043139.39-612141.8&-23.83&5.01&2.83&1.19&-0.58&0.91&67.91412&-61.36161 \\
J043140.34-611635.3&-&-&2.57&0.99&1.62&0.73&67.91808&-61.27647 \\
J043142.26-613139.2&0&0.32&0.09&0.18&0.67&0.46&67.92608&-61.52756 \\
J043142.88-613020.1&-&-&1.81&0.87&1.32&0.75&67.92867&-61.50558 \\
J043147.43-611754.8&-18.34&2.78&5.71&1.39&-17.05&2.94&67.94762&-61.29856 \\
J043148.16-611802.1&-&-&5.93&1.8&3.92&1.72&67.95067&-61.30058 \\
J043148.33-611633.8&-&-&8.23&1.52&1.53&1.39&67.95137&-61.27606 \\
J043150.03-611929.7&-10.85&4.49&7.57&1.73&3.58&1.43&67.95846&-61.32492 \\
J043150.25-613046.3&-&-&0.55&0.44&0.94&0.72&67.95937&-61.51286 \\
J043151.90-613215.1&-&-&-&-&-3.58&3.32&67.96625&-61.53753 \\
J043153.18-611850.0&-&-&-&-&8.37&2.34&67.97158&-61.31389 \\
J043153.46-612941.2&-&-&5.97&1.54&2.36&1.1&67.97275&-61.49478 \\
J043153.58-612738.5&-&-&6.31&1.58&0.78&1.18&67.97325&-61.46069 \\
J043153.72-612614.4&-&-&-&-&2.32&0.86&67.97383&-61.43733 \\
J043155.17-613042.0&-63.03&13.55&-1.44&0.74&-15.67&3.41&67.97987&-61.51167 \\
J043158.70-611624.1&-8.26&3.93&1.63&0.69&1.63&0.67&67.99458&-61.27336 \\
J043200.75-611503.0&-&-&2.8&0.97&1.23&0.78&68.00312&-61.25083 \\
J043202.49-611946.6&-&-&1.98&0.87&1.68&0.81&68.01037&-61.32961 \\
J043212.83-611232.9&-&-&2.56&0.94&2.21&0.93&68.05346&-61.20914 \\
J043213.28-612540.2&-5.97&3.25&3.8&1.38&2.81&1.03&68.05533&-61.42783 \\
J043213.90-613812.0&0&0.02&3&1.09&4.24&1.21&68.05792&-61.63667 \\
J043218.84-611920.3&-&-&5.01&1.35&4&1.29&68.0785&-61.32231 \\
J043221.93-611041.2&-6.94&6.72&-&-&1.82&0.88&68.09137&-61.17811 \\
J043222.39-613551.0&-117.41&86.44&-0.55&0.77&3.47&1.1&68.09329&-61.5975 \\
J043223.53-613340.0&-&-&0.34&0.33&1.58&0.69&68.09804&-61.56111 \\
J043225.62-612320.0&-31.43&6.95&-&-&4.65&2.05&68.10675&-61.38889 \\
J043226.65-612134.2&-&-&2.81&0.94&4.26&2.01&68.11104&-61.3595 \\
J043229.88-613044.6&0&1.09&1.21&0.56&1.58&0.72&68.1245&-61.51239 \\
J043233.84-613838.8&-&-&1.33&0.68&2.69&0.96&68.141&-61.64411 \\
J043234.08-611359.1&-4.32&1.46&1.59&0.75&-1.46&0.73&68.142&-61.23308 \\
J043234.77-611455.5&-&-&3.33&1.06&2.4&1.07&68.14487&-61.24875 \\
J043235.35-612945.4&-&-&3.93&1.3&1.51&0.96&68.14729&-61.49594 \\
J043239.62-612555.1&-&-&-&-&-4.8&2.15&68.16508&-61.43197 \\
J043240.10-612542.5&-&-&1.21&0.62&1.67&0.74&68.16708&-61.42847 \\
J043241.58-612303.8&-4.13&2.75&2.44&1.11&3.7&1.19&68.17325&-61.38439 \\
J043242.82-612811.1&-&-&2.97&1.01&0.84&0.75&68.17842&-61.46975 \\
J043244.94-611627.0&-13.95&3.89&5.41&1.2&1.14&0.79&68.18725&-61.27417 \\
J043245.88-613256.3&-&-&-&-&2.55&1.14&68.19117&-61.54897 \\
J043247.12-612107.2&0&0.97&3.64&1.06&3.12&1.4&68.19633&-61.352 \\
J043247.60-612511.3&-&-&-&-&6.04&1.33&68.19833&-61.41981 \\
J043248.45-611947.8&-&-&-&-&6.26&1.74&68.20187&-61.32994 \\
J043252.93-612936.9&-28.85&4.2&-&-&-12.75&3.68&68.22054&-61.49358 \\
J043254.13-612132.4&-&-&-&-&0.83&0.6&68.22554&-61.359 \\
J043254.18-612547.7&-&-&1.8&0.73&2.22&0.95&68.22575&-61.42992 \\
J043255.41-611910.4&0&0.25&1.68&0.7&4.01&1.06&68.23087&-61.31956 \\
J043257.83-612547.2&-&-&1.86&0.84&2.21&0.93&68.24096&-61.42978 \\
J043258.46-613005.1&0&0.5&1.59&0.73&2.48&0.83&68.24358&-61.50142 \\
J043306.66-612613.7&0&0.14&0.89&0.53&1.16&0.63&68.27775&-61.43714 \\
J043308.58-612115.4&-2.47&2.02&3.52&1.14&2.34&0.9&68.28575&-61.35428 \\
J043309.01-612312.1&-&-&6.9&1.57&-3.14&1.81&68.28754&-61.38669 \\
J043314.12-612034.0&-&-&2.23&0.87&1.78&0.73&68.30883&-61.34278 \\
J043315.31-612143.0&-&-&-&-&2.04&0.79&68.31379&-61.36194 \\
J043316.32-611800.9&-4.96&2.81&1.48&0.75&1.81&0.73&68.318&-61.30025 \\ 
J043322.76-612845.8&-&-&-&-&0.43&0.4&68.34483&-61.47939 \\ 
\enddata
\end{deluxetable}

\startlongtable
\begin{deluxetable}{cccccccccc}
\renewcommand\thetable{2}
	\tablecaption{Region Dimensions}
	\tabletypesize{\scriptsize}
	\tablecolumns{10}
	\tablenum{2}
	\tablewidth{0pt}
	\tablehead{
		\colhead{} &
		\colhead{RA} &
		\colhead{} & \colhead{} & \colhead{Dec.} & \colhead{} & \colhead{Off-axis} & \colhead{PSF Width} & \colhead{Inner Radius} & \colhead{Outer Radius}\\
		\colhead{hour} & \colhead{min} & \colhead{sec} & \colhead{degree} & \colhead{arcmin} & \colhead{arcsec} & \colhead{Radius (arcmin)} & \colhead{(arcsec)} & \colhead{(arcsec)} & \colhead{(arcsec)}
	}
	\startdata
4 &  31 &  50.03 &  -61 &  19 &  29.7 &  12.4337 &  19.9201 &  44.9201 &  69.9201 \\
4 &  30 &  34.31 &  -61 &  19 &  12.6 &  11.2017 &  17.973 &  42.973 &  67.973 \\
4 &  30 &  50.16 &  -61 &  15 &  43.9 &  14.6472 &  23.8213 &  48.8213 &  73.8213 \\
4 &  31 &  40.34 &  -61 &  16 &  35.3 &  14.724 &  23.9649 &  48.9649 &  73.9649 \\
4 &  30 &  59.59 &  -61 &  25 &  7.2 &  4.07797 &  11.037 &  36.037 &  61.037 \\
4 &  30 &  8.03 &  -61 &  18 &  9.3 &  13.7105 &  22.112 &  47.112 &  72.112 \\
4 &  31 &  39.39 &  -61 &  21 &  41.8 &  9.58788 &  15.7072 &  40.7072 &  65.7072 \\
4 &  31 &  13.2 &  -61 &  10 &  37.9 &  20.4041 &  35.8627 &  60.8627 &  85.8627 \\
4 &  32 &  47.12 &  -61 &  21 &  7.2 &  16.8322 &  28.1034 &  53.1034 &  78.1034 \\
4 &  33 &  14.12 &  -61 &  20 &  34 &  20.3085 &  35.644 &  60.644 &  85.644 \\
4 &  32 &  55.41 &  -61 &  19 &  10.4 &  18.9509 &  32.6022 &  57.6022 &  82.6022 \\
4 &  32 &  26.65 &  -61 &  21 &  34.2 &  14.2444 &  23.0761 &  48.0761 &  73.0761 \\
4 &  30 &  28.36 &  -61 &  17 &  4 &  13.7331 &  22.1522 &  47.1522 &  72.1522 \\
4 &  29 &  44.51 &  -61 &  14 &  56.3 &  18.4374 &  31.4832 &  56.4832 &  81.4832 \\
4 &  29 &  57.45 &  -61 &  11 &  45.7 &  20.767 &  36.6974 &  61.6974 &  86.6974 \\
4 &  30 &  56.87 &  -61 &  13 &  33.5 &  17.0372 &  28.5244 &  53.5244 &  78.5244 \\
4 &  30 &  58.75 &  -61 &  10 &  21 &  20.6304 &  36.3822 &  61.3822 &  86.3822 \\
4 &  30 &  51.24 &  -61 &  10 &  59.7 &  19.9349 &  34.7953 &  59.7953 &  84.7953 \\
4 &  29 &  39.59 &  -61 &  10 &  34.5 &  23.0053 &  42.0159 &  67.0159 &  92.0159 \\
4 &  30 &  50.35 &  -61 &  14 &  6 &  16.4688 &  27.3644 &  52.3644 &  77.3644 \\
4 &  29 &  5.27 &  -61 &  13 &  2.4 &  23.3189 &  42.7833 &  67.7833 &  92.7833 \\
4 &  30 &  4.47 &  -61 &  23 &  25.7 &  9.43714 &  15.5137 &  40.5137 &  65.5137 \\
4 &  30 &  45.38 &  -61 &  23 &  35.7 &  6.06744 &  12.1249 &  37.1249 &  62.1249 \\
4 &  29 &  18.62 &  -61 &  19 &  29.8 &  16.9995 &  28.4468 &  53.4468 &  78.4468 \\
4 &  29 &  47.62 &  -61 &  23 &  37 &  11.1649 &  17.9177 &  42.9177 &  67.9177 \\
4 &  30 &  22.67 &  -61 &  23 &  19.6 &  7.79429 &  13.6278 &  38.6278 &  63.6278 \\
4 &  29 &  24.75 &  -61 &  19 &  5.7 &  16.6465 &  27.7244 &  52.7244 &  77.7244 \\
4 &  30 &  53.08 &  -61 &  25 &  14.2 &  4.02546 &  11.0173 &  36.0173 &  61.0173 \\
4 &  30 &  28.68 &  -61 &  19 &  39.8 &  10.9741 &  17.6329 &  42.6329 &  67.6329 \\
4 &  29 &  32.77 &  -61 &  14 &  52.5 &  19.3951 &  33.5845 &  58.5845 &  83.5845 \\
4 &  29 &  27.18 &  -61 &  16 &  51.9 &  18.1529 &  30.8708 &  55.8708 &  80.8708 \\
4 &  31 &  8.71 &  -61 &  27 &  7 &  2.25179 &  10.5845 &  35.5845 &  60.5845 \\
4 &  29 &  2.33 &  -61 &  24 &  17.2 &  16.4114 &  27.2487 &  52.2487 &  77.2487 \\
4 &  29 &  34.78 &  -61 &  28 &  3.7 &  11.2969 &  18.1171 &  43.1171 &  68.1171 \\
4 &  29 &  33.1 &  -61 &  15 &  36.3 &  18.7164 &  32.089 &  57.089 &  82.089 \\
4 &  30 &  20.03 &  -61 &  26 &  9.7 &  5.97435 &  12.0586 &  37.0586 &  62.0586 \\
4 &  29 &  24.92 &  -61 &  26 &  3.4 &  12.9547 &  20.7942 &  45.7942 &  70.7942 \\
4 &  29 &  25.36 &  -61 &  33 &  45.9 &  13.712 &  22.1146 &  47.1146 &  72.1146 \\
4 &  30 &  37.69 &  -61 &  29 &  2.7 &  2.86953 &  10.6878 &  35.6878 &  60.6878 \\
4 &  30 &  14.66 &  -61 &  28 &  8.7 &  5.97325 &  12.0578 &  37.0578 &  62.0578 \\
4 &  30 &  57 &  -61 &  31 &  7.3 &  2.66258 &  10.6481 &  35.6481 &  60.6481 \\
4 &  29 &  48.73 &  -61 &  37 &  32.2 &  13.593 &  21.9034 &  46.9034 &  71.9034 \\
4 &  30 &  52.25 &  -61 &  28 &  13.8 &  1.08041 &  10.4926 &  35.4926 &  60.4926 \\
4 &  29 &  8.67 &  -61 &  39 &  6.4 &  18.74 &  32.1406 &  57.1406 &  82.1406 \\
4 &  30 &  32.85 &  -61 &  37 &  55.5 &  10.8505 &  17.4511 &  42.4511 &  67.4511 \\
4 &  30 &  45.32 &  -61 &  33 &  0.4 &  5.10075 &  11.5112 &  36.5112 &  61.5112 \\
4 &  30 &  39.53 &  -61 &  35 &  7.8 &  7.60024 &  13.4335 &  38.4335 &  63.4335 \\
4 &  30 &  18.11 &  -61 &  35 &  8.8 &  9.00532 &  14.9778 &  39.9778 &  64.9778 \\
4 &  30 &  37.57 &  -61 &  42 &  49.6 &  16.0239 &  26.474 &  51.474 &  76.474 \\
4 &  30 &  59.77 &  -61 &  34 &  38.5 &  6.5948 &  12.529 &  37.529 &  62.529 \\
4 &  31 &  31.08 &  -61 &  42 &  8.8 &  15.6023 &  25.6447 &  50.6447 &  75.6447 \\
4 &  31 &  17.25 &  -61 &  31 &  49.8 &  4.21908 &  11.0922 &  36.0922 &  61.0922 \\
4 &  31 &  42.26 &  -61 &  31 &  39.2 &  6.63035 &  12.558 &  37.558 &  62.558 \\
4 &  32 &  13.9 &  -61 &  38 &  12 &  14.5553 &  23.65 &  48.65 &  73.65 \\
4 &  32 &  22.39 &  -61 &  35 &  51 &  13.6799 &  22.0574 &  47.0574 &  72.0574 \\
4 &  32 &  33.84 &  -61 &  38 &  38.8 &  16.8205 &  28.0793 &  53.0793 &  78.0793 \\
4 &  32 &  35.35 &  -61 &  29 &  45.4 &  12.934 &  20.759 &  45.759 &  70.759 \\
4 &  32 &  40.1 &  -61 &  25 &  42.5 &  13.9603 &  22.5597 &  47.5597 &  72.5597 \\
4 &  31 &  55.17 &  -61 &  30 &  42 &  7.81878 &  13.6528 &  38.6528 &  63.6528 \\
4 &  32 &  45.88 &  -61 &  32 &  56.3 &  15.0305 &  24.5435 &  49.5435 &  74.5435 \\
4 &  32 &  42.82 &  -61 &  28 &  11.1 &  13.9052 &  22.4605 &  47.4605 &  72.4605 \\
4 &  31 &  23.3 &  -61 &  27 &  26.9 &  3.56717 &  10.8634 &  35.8634 &  60.8634 \\
4 &  32 &  47.6 &  -61 &  25 &  11.3 &  15.0819 &  24.6414 &  49.6414 &  74.6414 \\
4 &  32 &  13.28 &  -61 &  25 &  40.2 &  10.5296 &  16.9877 &  41.9877 &  66.9877 \\
4 &  32 &  41.58 &  -61 &  23 &  3.8 &  15.1529 &  24.7769 &  49.7769 &  74.7769 \\
4 &  33 &  9.01 &  -61 &  23 &  12.1 &  18.4975 &  31.6132 &  56.6132 &  81.6132 \\
4 &  31 &  37.44 &  -61 &  25 &  52.9 &  6.07204 &  12.1282 &  37.1282 &  62.1282 \\
4 &  33 &  16.32 &  -61 &  18 &  0.9 &  22.008 &  39.6109 &  64.6109 &  89.6109 \\
4 &  31 &  14.59 &  -61 &  23 &  44.7 &  5.99355 &  12.0722 &  37.0722 &  62.0722 \\
4 &  31 &  21.86 &  -61 &  25 &  17.7 &  4.94247 &  11.4263 &  36.4263 &  61.4263 \\
4 &  33 &  6.66 &  -61 &  26 &  13.7 &  17.3227 &  29.1164 &  54.1164 &  79.1164 \\
4 &  32 &  54.18 &  -61 &  25 &  47.7 &  15.7737 &  25.9801 &  50.9801 &  75.9801 \\
4 &  33 &  15.31 &  -61 &  21 &  43 &  19.9006 &  34.7178 &  59.7178 &  84.7178 \\
4 &  31 &  58.7 &  -61 &  16 &  24.1 &  15.999 &  26.4245 &  51.4245 &  76.4245 \\
4 &  32 &  34.77 &  -61 &  14 &  55.5 &  20.1417 &  35.2641 &  60.2641 &  85.2641 \\
4 &  32 &  34.08 &  -61 &  13 &  59.1 &  20.9086 &  37.0253 &  62.0253 &  87.0253 \\
4 &  30 &  51.87 &  -61 &  23 &  24.4 &  6.07313 &  12.129 &  37.129 &  62.129 \\
4 &  32 &  44.94 &  -61 &  16 &  27 &  19.81 &  34.5136 &  59.5136 &  84.5136 \\
4 &  30 &  41.79 &  -61 &  22 &  43.1 &  7.15115 &  13.0081 &  38.0081 &  63.0081 \\
4 &  32 &  12.83 &  -61 &  12 &  32.9 &  20.6981 &  36.5383 &  61.5383 &  86.5383 \\
4 &  30 &  43.25 &  -61 &  21 &  6.9 &  8.82232 &  14.7591 &  39.7591 &  64.7591 \\
4 &  30 &  54.21 &  -61 &  21 &  3.2 &  8.65877 &  14.568 &  39.568 &  64.568 \\
4 &  31 &  8.81 &  -61 &  25 &  10.1 &  4.23185 &  11.0973 &  36.0973 &  61.0973 \\
4 &  32 &  0.75 &  -61 &  15 &  3 &  17.4571 &  29.3971 &  54.3971 &  79.3971 \\
4 &  30 &  17.05 &  -61 &  18 &  44.1 &  12.5641 &  20.1361 &  45.1361 &  70.1361 \\
4 &  29 &  51.03 &  -61 &  11 &  27.3 &  21.4363 &  38.2574 &  63.2574 &  88.2574 \\
4 &  30 &  41.88 &  -61 &  17 &  9.9 &  13.1943 &  21.2058 &  46.2058 &  71.2058 \\
4 &  30 &  44.89 &  -61 &  11 &  4.6 &  19.9066 &  34.7315 &  59.7315 &  84.7315 \\
4 &  30 &  41.86 &  -61 &  13 &  3 &  17.7545 &  30.0227 &  55.0227 &  80.0227 \\
4 &  29 &  37.98 &  -61 &  14 &  45.5 &  19.0932 &  32.9156 &  57.9156 &  82.9156 \\
4 &  29 &  32.16 &  -61 &  13 &  13 &  20.9638 &  37.1533 &  62.1533 &  87.1533 \\
4 &  29 &  26.28 &  -61 &  11 &  55.9 &  22.6027 &  41.0383 &  66.0383 &  91.0383 \\
4 &  29 &  2.02 &  -61 &  13 &  54.9 &  22.8824 &  41.7164 &  66.7164 &  91.7164 \\
4 &  30 &  7.6 &  -61 &  20 &  3.6 &  11.939 &  19.1174 &  44.1174 &  69.1174 \\
4 &  30 &  33.1 &  -61 &  24 &  22.6 &  6.01252 &  12.0856 &  37.0856 &  62.0856 \\
4 &  29 &  7.2 &  -61 &  20 &  18.3 &  17.7337 &  29.9788 &  54.9788 &  79.9788 \\
4 &  29 &  12.95 &  -61 &  15 &  8.4 &  20.8735 &  36.9438 &  61.9438 &  86.9438 \\
4 &  30 &  59.55 &  -61 &  27 &  40.1 &  1.22339 &  10.4981 &  35.4981 &  60.4981 \\
4 &  29 &  51.96 &  -61 &  18 &  12.9 &  14.8499 &  24.2016 &  49.2016 &  74.2016 \\
4 &  29 &  27.89 &  -61 &  23 &  37 &  13.498 &  21.7356 &  46.7356 &  71.7356 \\
4 &  30 &  3.09 &  -61 &  17 &  14 &  14.9374 &  24.367 &  49.367 &  74.367 \\
4 &  30 &  30.43 &  -61 &  20 &  23.1 &  10.1359 &  16.4377 &  41.4377 &  66.4377 \\
4 &  30 &  41.87 &  -61 &  27 &  15.8 &  2.84001 &  10.6818 &  35.6818 &  60.6818 \\
4 &  29 &  17.68 &  -61 &  28 &  18.2 &  13.5667 &  21.8568 &  46.8568 &  71.8568 \\
4 &  30 &  33.23 &  -61 &  33 &  0.7 &  5.88135 &  11.994 &  36.994 &  61.994 \\
4 &  29 &  35.69 &  -61 &  31 &  40.7 &  11.6083 &  18.5961 &  43.5961 &  68.5961 \\
4 &  30 &  21.56 &  -61 &  28 &  48.7 &  5.00984 &  11.4619 &  36.4619 &  61.4619 \\
4 &  28 &  47.65 &  -61 &  30 &  52.9 &  17.7234 &  29.9569 &  54.9569 &  79.9569 \\
4 &  30 &  53.73 &  -61 &  30 &  5.7 &  1.65623 &  10.5234 &  35.5234 &  60.5234 \\
4 &  30 &  26.74 &  -61 &  30 &  31.9 &  4.75041 &  11.329 &  36.329 &  61.329 \\
4 &  28 &  56.57 &  -61 &  37 &  53.8 &  19.284 &  33.3377 &  58.3377 &  83.3377 \\
4 &  31 &  4.61 &  -61 &  28 &  2.6 &  1.09718 &  10.4932 &  35.4932 &  60.4932 \\
4 &  30 &  28.84 &  -61 &  39 &  32.5 &  12.7379 &  20.427 &  45.427 &  70.427 \\
4 &  30 &  26.85 &  -61 &  35 &  3.1 &  8.25806 &  14.1173 &  39.1173 &  64.1173 \\
4 &  31 &  12.64 &  -61 &  27 &  28.9 &  2.31846 &  10.5935 &  35.5935 &  60.5935 \\
4 &  31 &  8.34 &  -61 &  34 &  13.8 &  6.25812 &  12.2653 &  37.2653 &  62.2653 \\
4 &  31 &  7.7 &  -61 &  30 &  12.1 &  1.99237 &  10.5537 &  35.5537 &  60.5537 \\
4 &  31 &  30.61 &  -61 &  38 &  45 &  11.9659 &  19.1603 &  44.1603 &  69.1603 \\
4 &  30 &  50.84 &  -61 &  35 &  11.7 &  7.29611 &  13.1417 &  38.1417 &  63.1417 \\
4 &  30 &  6.99 &  -61 &  35 &  51.2 &  10.5581 &  17.0283 &  42.0283 &  67.0283 \\
4 &  31 &  39.3 &  -61 &  39 &  51.6 &  13.5535 &  21.8335 &  46.8335 &  71.8335 \\
4 &  31 &  21.54 &  -61 &  30 &  16.3 &  3.45809 &  10.8315 &  35.8315 &  60.8315 \\
4 &  31 &  17.87 &  -61 &  35 &  37.9 &  8.10492 &  13.9518 &  38.9518 &  63.9518 \\
4 &  31 &  42.88 &  -61 &  30 &  20.1 &  6.12744 &  12.1684 &  37.1684 &  62.1684 \\
4 &  32 &  29.88 &  -61 &  30 &  44.6 &  12.3527 &  19.7869 &  44.7869 &  69.7869 \\
4 &  31 &  53.46 &  -61 &  29 &  41.2 &  7.35707 &  13.1989 &  38.1989 &  63.1989 \\
4 &  32 &  4.87 &  -61 &  28 &  49.9 &  8.81154 &  14.7464 &  39.7464 &  64.7464 \\
4 &  32 &  23.53 &  -61 &  33 &  40 &  12.5615 &  20.1319 &  45.1319 &  70.1319 \\
4 &  32 &  58.46 &  -61 &  30 &  5.1 &  16.044 &  26.5138 &  51.5138 &  76.5138 \\
4 &  33 &  22.76 &  -61 &  28 &  45.8 &  19.2314 &  33.2211 &  58.2211 &  83.2211 \\
4 &  31 &  53.72 &  -61 &  26 &  14.4 &  7.84943 &  13.6842 &  38.6842 &  63.6842 \\
4 &  31 &  53.58 &  -61 &  27 &  38.5 &  7.40951 &  13.2487 &  38.2487 &  63.2487 \\
4 &  32 &  57.83 &  -61 &  25 &  47.2 &  16.2537 &  26.9321 &  51.9321 &  76.9321 \\
4 &  33 &  8.58 &  -61 &  21 &  15.4 &  19.2968 &  33.366 &  58.366 &  83.366 \\
4 &  32 &  54.13 &  -61 &  21 &  32.4 &  17.4229 &  29.3254 &  54.3254 &  79.3254 \\
4 &  31 &  53.18 &  -61 &  18 &  50 &  13.2862 &  21.3651 &  46.3651 &  71.3651 \\
4 &  31 &  48.33 &  -61 &  16 &  33.8 &  15.1859 &  24.84 &  49.84 &  74.84 \\
4 &  30 &  56.9 &  -61 &  21 &  47.4 &  7.81436 &  13.6483 &  38.6483 &  63.6483 \\
4 &  31 &  9.75 &  -61 &  25 &  35.8 &  3.82392 &  10.9457 &  35.9457 &  60.9457 \\
4 &  30 &  32.53 &  -61 &  12 &  57.5 &  18.0609 &  30.6739 &  55.6739 &  80.6739 \\
4 &  30 &  39.66 &  -61 &  17 &  51.9 &  12.481 &  19.9983 &  44.9983 &  69.9983 \\
4 &  29 &  5.22 &  -61 &  13 &  21.7 &  23.0514 &  42.1282 &  67.1282 &  92.1282 \\
4 &  29 &  18.93 &  -61 &  14 &  19.3 &  21.0332 &  37.3147 &  62.3147 &  87.3147 \\
4 &  29 &  36.13 &  -61 &  14 &  50.2 &  19.1642 &  33.0723 &  58.0723 &  83.0723 \\
4 &  30 &  16.68 &  -61 &  22 &  39.4 &  8.88398 &  14.8322 &  39.8322 &  64.8322 \\
4 &  29 &  16.95 &  -61 &  19 &  53.1 &  16.9168 &  28.2769 &  53.2769 &  78.2769 \\
4 &  30 &  12.3 &  -61 &  19 &  46.4 &  11.8584 &  18.9892 &  43.9892 &  68.9892 \\
4 &  30 &  8.45 &  -61 &  25 &  29.8 &  7.69306 &  13.5257 &  38.5257 &  63.5257 \\
4 &  31 &  13.24 &  -61 &  27 &  12.2 &  2.58166 &  10.6341 &  35.6341 &  60.6341 \\
4 &  30 &  0.74 &  -61 &  25 &  54.3 &  8.43068 &  14.3083 &  39.3083 &  64.3083 \\
4 &  30 &  25.99 &  -61 &  27 &  45.9 &  4.55609 &  11.2369 &  36.2369 &  61.2369 \\
4 &  30 &  45.64 &  -61 &  35 &  38.7 &  7.92215 &  13.7593 &  38.7593 &  63.7593 \\
4 &  30 &  58.25 &  -61 &  35 &  36.2 &  7.67245 &  13.5051 &  38.5051 &  63.5051 \\
4 &  31 &  51.9 &  -61 &  32 &  15.1 &  8.08144 &  13.9268 &  38.9268 &  63.9268 \\
4 &  31 &  17.66 &  -61 &  27 &  43.7 &  2.75006 &  10.6643 &  35.6643 &  60.6643 \\
4 &  32 &  52.93 &  -61 &  29 &  36.9 &  15.2675 &  24.9967 &  49.9967 &  74.9967 \\
4 &  32 &  25.62 &  -61 &  23 &  20 &  13.1002 &  21.0435 &  46.0435 &  71.0435 \\
4 &  32 &  48.45 &  -61 &  19 &  47.8 &  17.7839 &  30.0849 &  55.0849 &  80.0849 \\
4 &  31 &  26.94 &  -61 &  23 &  1.3 &  7.43901 &  13.2768 &  38.2768 &  63.2768 \\
4 &  32 &  18.84 &  -61 &  19 &  20.3 &  15.0363 &  24.5546 &  49.5546 &  74.5546 \\
4 &  31 &  48.16 &  -61 &  18 &  2.1 &  13.7081 &  22.1077 &  47.1077 &  72.1077 \\
4 &  32 &  2.49 &  -61 &  19 &  46.6 &  13.1831 &  21.1864 &  46.1864 &  71.1864 \\
4 &  32 &  21.93 &  -61 &  10 &  41.2 &  23.1181 &  42.2911 &  67.2911 &  92.2911 \\
4 &  29 &  33.01 &  -61 &  11 &  24.4 &  22.6216 &  41.084 &  66.084 &  91.084 \\
4 &  30 &  46.27 &  -61 &  11 &  32.2 &  19.3762 &  33.5426 &  58.5426 &  83.5426 \\
4 &  30 &  52.13 &  -61 &  10 &  34.7 &  20.3955 &  35.8428 &  60.8428 &  85.8428 \\
4 &  30 &  18.94 &  -61 &  18 &  6.9 &  13.0832 &  21.0142 &  46.0142 &  71.0142 \\
4 &  30 &  3.77 &  -61 &  15 &  55.4 &  16.1829 &  26.7904 &  51.7904 &  76.7904 \\
4 &  30 &  4.38 &  -61 &  22 &  55.5 &  9.81363 &  16.0031 &  41.0031 &  66.0031 \\
4 &  30 &  55.49 &  -61 &  20 &  14.3 &  9.55989 &  15.671 &  40.671 &  65.671 \\
4 &  30 &  38.92 &  -61 &  28 &  37.5 &  2.69135 &  10.6533 &  35.6533 &  60.6533 \\
4 &  29 &  44.77 &  -61 &  30 &  54.9 &  10.2157 &  16.5475 &  41.5475 &  66.5475 \\
4 &  31 &  0.91 &  -61 &  29 &  57.9 &  1.37648 &  10.5055 &  35.5055 &  60.5055 \\
4 &  30 &  14.72 &  -61 &  31 &  45.4 &  6.80727 &  12.7056 &  37.7056 &  62.7056 \\
4 &  30 &  32.07 &  -61 &  42 &  0.6 &  15.2798 &  25.0203 &  50.0203 &  75.0203 \\
4 &  30 &  56.84 &  -61 &  32 &  2.3 &  3.68776 &  10.9008 &  35.9008 &  60.9008 \\
4 &  31 &  4.72 &  -61 &  30 &  53.4 &  2.50857 &  10.622 &  35.622 &  60.622 \\
4 &  31 &  50.25 &  -61 &  30 &  46.3 &  7.21385 &  13.0654 &  38.0654 &  63.0654 \\
	\enddata
\end{deluxetable}

\startlongtable
\begin{deluxetable}{cccccccccc}
    \tabletypesize{\scriptsize}
	\tablecolumns{10}
	\tablenum{3}
	\tablewidth{0pt}
	\tablehead{ 
		\colhead{} &
		\colhead{RA} &
		\colhead{} & \colhead{} & \colhead{Dec.} & \colhead{} & \colhead{Count Rate} & \colhead{Error} & \colhead{Flux} & \colhead{Norm} \\
		\colhead{hour} & \colhead{min.} & \colhead{sec.} & \colhead{degree} & \colhead{arcmin.} & \colhead{arcsec.} & \colhead{(cts/s)} & \colhead{} & \colhead{(ergs/cm$^2$/s)} & \colhead{} 
	}
\startdata
4 & 31 & 50.03 & -61 & 19 & 29.7 & 1.634E-02 & 1.126E-03 & 2.257E-13 & 3.924E-04 \\
4 & 30 & 34.31 & -61 & 19 & 12.6 & 9.145E-03 & 8.488E-04 & 1.264E-13 & 2.087E-04   \\
4 & 30 & 50.16 & -61 & 15 & 43.9 & 4.326E-03 & 5.980E-04 & 5.977E-14 & 9.871E-05  \\
4 & 31 & 40.34 & -61 & 16 & 35.3 & 8.956E-03 & 8.421E-04 & 1.237E-13 & 2.044E-04  \\
4 & 30 & 59.59 & -61 & 25 & 7.2 & 2.444E-02  & 1.371E-03 & 3.377E-13 & 5.577E-04  \\
4 & 30 & 8.03  & -61 & 18 & 9.3 & 5.020E-03  & 6.399E-04 & 6.936E-14 & 1.146E-04  \\
4 & 31 & 39.39 & -61 & 21 & 41.8 & 2.829E-02 & 1.474E-03 & 3.909E-13 & 6.455E-04  \\
4 & 31 & 13.2  & -61 & 10 & 37.9 & 1.444E-03 & 3.803E-04 & 1.995E-14 & 3.295E-05  \\
4 & 32 & 47.12 & -61 & 21 & 7.2 & 6.577E-03  & 7.284E-04 & 9.083E-14 & 1.579E-04  \\
4 & 33 & 14.12 & -61 & 20 & 34 & 1.521E-03   & 3.878E-04 & 2.102E-14 & 3.471E-05  \\
4 & 32 & 55.41 & -61 & 19 & 10.4 & 3.589E-03 & 5.531E-04 & 4.957E-14 & 8.618E-05  \\
4 & 32 & 26.65 & -61 & 21 & 34.2 & 1.048E-02 & 9.079E-04 & 1.448E-13 & 2.391E-04  \\
4 & 30 & 28.36 & -61 & 17 & 4.0 & 4.716E-03    & 6.216E-04 & 6.516E-14 & 1.076E-04  \\
4 & 29 & 44.51 & -61 & 14 & 56.3 & 2.154E-03 & 4.432E-04 & 2.976E-14 & 4.915E-05  \\
4 & 29 & 57.45 & -61 & 11 & 45.7 & 1.970E-03 & 4.301E-04 & 2.722E-14 & 4.495E-05  \\
4 & 30 & 56.87 & -61 & 13 & 33.5 & 1.945E-03 & 4.232E-04 & 2.687E-14 & 4.438E-05 \\
4 & 30 & 58.75 & -61 & 10 & 21 & 1.440E-03   & 3.803E-04 & 1.990E-14 & 3.286E-05  \\
4 & 30 & 51.24 & -61 & 10 & 59.7 & 1.982E-03 & 4.301E-04 & 2.739E-14 & 4.523E-05  \\
4 & 29 & 39.59 & -61 & 10 & 34.5 & 1.026E-03 & 3.405E-04 & 1.418E-14 & 2.341E-05 \\
4 & 30 & 50.35 & -61 & 14 & 6 & 2.787E-03    & 4.924E-04 & 3.851E-14 & 6.360E-05  \\
4 & 29 & 5.27  & -61 & 13 & 2.4 & 1.553E-03  & 3.953E-04 & 2.146E-14 & 3.544E-05 \\
4 & 30 & 4.47  & -61 & 23 & 25.7 & 7.186E-03 & 7.554E-04 & 9.929E-14 & 1.640E-04  \\
4 & 30 & 45.38 & -61 & 23 & 35.7 & 1.480E-02 & 1.071E-03 & 2.044E-13 & 3.554E-04 \\
4 & 29 & 18.62 & -61 & 19 & 29.8 & 1.490E-03 & 3.802E-04 & 2.059E-14 & 3.400E-05 \\
4 & 29 & 47.62 & -61 & 23 & 37 & 3.907E-03   & 5.683E-04 & 5.398E-14 & 8.915E-05 \\
4 & 30 & 22.67 & -61 & 23 & 19.6 & 1.448E-02 & 1.060E-03 & 2.001E-13 & 3.304E-04  \\
4 & 29 & 24.75 & -61 & 19 & 5.7 & 1.342E-03  & 3.647E-04 & 1.854E-14 & 3.062E-05  \\
4 & 30 & 53.08 & -61 & 25 & 14.2 & 2.126E-02 & 1.279E-03 & 2.938E-13 & 4.851E-04  \\
4 & 30 & 28.68 & -61 & 19 & 39.8 & 8.539E-03 & 8.212E-04 & 1.179E-13 & 2.050E-04  \\
4 & 29 & 32.77 & -61 & 14 & 52.5 & 1.329E-03 & 3.565E-04 & 1.836E-14 & 3.033E-05  \\
4 & 29 & 27.18 & -61 & 16 & 51.9 & 1.778E-03 & 4.094E-04 & 2.457E-14 & 4.057E-05  \\
4 & 31 & 8.71  & -61 & 27 & 7 & 5.321E-02    & 2.015E-03 & 7.349E-13 & 1.278E-03  \\
4 & 29 & 2.33  & -61 & 24 & 17.2 & 1.725E-03 & 4.022E-04 & 2.383E-14 & 3.936E-05 \\
4 & 29 & 34.78 & -61 & 28 & 3.7 & 3.299E-03  & 5.262E-04 & 4.558E-14 & 7.528E-05  \\
4 & 29 & 33.1  & -61 & 15 & 36.3 & 2.226E-03  & 4.497E-04 & 3.076E-14 & 5.079E-05  \\
4 & 30 & 20.03 & -61 & 26 & 9.7 & 1.032E-02  & 8.982E-04 & 1.426E-13 & 2.355E-04  \\
4 & 29 & 24.92 & -61 & 26 & 3.4 & 2.372E-03  & 4.558E-04 & 3.277E-14 & 5.413E-05  \\
4 & 29 & 25.36 & -61 & 33 & 45.9 & 1.225E-03 & 3.484E-04 & 1.693E-14 & 2.795E-05 \\
4 & 30 & 37.69 & -61 & 29 & 2.7 & 1.086E-02  & 9.204E-04 & 1.501E-13 & 2.478E-04  \\
4 & 30 & 14.66 & -61 & 28 & 8.7 & 9.561E-03  & 8.656E-04 & 1.321E-13 & 2.182E-04 \\
4 & 30 & 57.00 & -61 & 31 & 7.3 & 8.431E-03 & 8.141E-04 & 1.165E-13 & 1.924E-04  \\
4 & 29 & 48.73 & -61 & 37 & 32.2 & 1.226E-03 & 3.484E-04 & 1.694E-14 & 2.798E-05  \\
4 & 30 & 52.25 & -61 & 28 & 13.8 & 1.883E-02 & 1.205E-03 & 2.601E-13 & 4.521E-04  \\
4 & 29 & 8.67  & -61 & 39 & 6.4 & 1.770E-03 & 4.094E-04 & 2.446E-14 & 4.039E-05  \\
4 & 30 & 32.85 & -61 & 37 & 55.5 & 1.785E-03 & 4.021E-04 & 2.466E-14 & 4.073E-05  \\
4 & 30 & 45.32 & -61 & 33 & 0.4 & 4.251E-03 & 5.882E-04 & 5.874E-14 & 9.700E-05  \\
4 & 30 & 39.53 & -61 & 35 & 7.8 & 2.569E-03 & 4.682E-04 & 3.550E-14 & 5.862E-05 \\
4 & 30 & 18.11 & -61 & 35 & 8.8 & 3.167E-03 & 5.151E-04 & 4.376E-14 & 7.227E-05 \\
4 & 30 & 37.57 & -61 & 42 & 49.6 & 1.350E-03 & 3.647E-04 & 1.865E-14 & 3.081E-05  \\
4 & 30 & 59.77 & -61 & 34 & 38.5 & 3.789E-03 & 5.580E-04 & 5.235E-14 & 8.646E-05  \\
4 & 31 & 31.08 & -61 & 42 & 8.8 & 5.201E-04 & 2.638E-04 & 7.186E-15 & 1.187E-05  \\
4 & 31 & 17.25 & -61 & 31 & 49.8 & 9.111E-03 & 8.454E-04 & 1.259E-13 & 2.079E-04 \\
4 & 31 & 42.26 & -61 & 31 & 39.2 & 8.115E-03 & 7.999E-04 & 1.121E-13 & 1.852E-04  \\
4 & 32 & 13.9  & -61 & 38 & 12 & 1.216E-03 & 3.484E-04 & 1.680E-14 & 2.775E-05  \\
4 & 32 & 22.39 & -61 & 35 & 51 & 1.984E-03 & 4.231E-04 & 2.741E-14 & 4.527E-05  \\
4 & 32 & 33.84 & -61 & 38 & 38.8 & 1.947E-03 & 4.232E-04 & 2.690E-14 & 4.443E-05  \\
4 & 32 & 35.35 & -61 & 29 & 45.4 & 5.484E-03 & 6.663E-04 & 7.577E-14 & 1.251E-04  \\
4 & 32 & 40.1  & -61 & 25 & 42.5 & 1.010E-02 & 8.919E-04 & 1.396E-13 & 2.305E-04  \\
4 & 31 & 55.17 & -61 & 30 & 42 & 8.868E-03 & 8.351E-04 & 1.225E-13 & 2.129E-04  \\
4 & 32 & 45.88 & -61 & 32 & 56.3 & 2.804E-03 & 4.924E-04 & 3.874E-14 & 6.398E-05  \\
4 & 32 & 42.82 & -61 & 28 & 11.1 & 5.853E-03 & 6.876E-04 & 8.087E-14 & 1.336E-04  \\
4 & 31 & 23.3  & -61 & 27 & 26.9 & 6.611E-02 & 2.245E-03 & 9.135E-13 & 1.509E-03  \\
4 & 32 & 47.6  & -61 & 25 & 11.3 & 6.143E-03 & 7.042E-04 & 8.488E-14 & 1.402E-04  \\
4 & 32 & 13.28 & -61 & 25 & 40.2 & 1.970E-02 & 1.233E-03 & 2.721E-13 & 4.730E-04 \\
4 & 32 & 41.58 & -61 & 23 & 3.8 & 9.254E-03 & 8.556E-04 & 1.279E-13 & 2.112E-04  \\
4 & 33 & 9.01  & -61 & 23 & 12.1 & 3.671E-03 & 5.583E-04 & 5.072E-14 & 8.377E-05  \\
4 & 31 & 37.44 & -61 & 25 & 52.9 & 4.432E-02 & 1.841E-03 & 6.121E-13 & 1.064E-03  \\
4 & 33 & 16.32 & -61 & 18 & 0.9 & 9.650E-04 & 3.319E-04 & 1.333E-14 & 2.202E-05  \\
4 & 31 & 14.59 & -61 & 23 & 44.7 & 3.370E-02 & 1.607E-03 & 4.654E-13 & 8.092E-04  \\
4 & 31 & 21.86 & -61 & 25 & 17.7 & 5.875E-02 & 2.117E-03 & 8.118E-13 & 1.341E-03 \\
4 & 33 & 6.66  & -61 & 26 & 13.7 & 2.396E-03 & 4.622E-04 & 3.309E-14 & 5.753E-05  \\
4 & 32 & 54.18 & -61 & 25 & 47.7 & 4.845E-03 & 6.309E-04 & 6.691E-14 & 1.163E-04 \\
4 & 33 & 15.31 & -61 & 21 & 43 & 2.665E-03 & 4.866E-04 & 3.681E-14 & 6.399E-05  \\
4 & 31 & 58.7  & -61 & 16 & 24.1 & 6.891E-03 & 7.440E-04 & 9.517E-14 & 1.655E-04 \\
4 & 32 & 34.77 & -61 & 14 & 55.5 & 2.510E-03 & 4.747E-04 & 3.466E-14 & 6.027E-05  \\
4 & 32 & 34.08 & -61 & 13 & 59.1 & 1.892E-03 & 4.234E-04 & 2.613E-14 & 4.543E-05 \\
4 & 30 & 51.87 & -61 & 23 & 24.4 & 1.487E-02 & 1.074E-03 & 2.054E-13 & 3.571E-04  \\
4 & 32 & 44.94 & -61 & 16 & 27 & 2.439E-03 & 4.685E-04 & 3.368E-14 & 5.856E-05  \\
4 & 30 & 41.79 & -61 & 22 & 43.1 & 1.350E-02 & 1.024E-03 & 1.864E-13 & 3.242E-04  \\
4 & 32 & 12.83 & -61 & 12 & 32.9 & 2.123E-03 & 4.433E-04 & 2.932E-14 & 5.098E-05  \\
4 & 30 & 43.25 & -61 & 21 & 6.9 & 1.266E-02 & 9.927E-04 & 1.748E-13 & 3.040E-04 \\
4 & 30 & 54.21 & -61 & 21 & 3.2 & 1.501E-02 & 1.079E-03 & 2.073E-13 & 3.604E-04 \\
4 & 31 & 8.81  & -61 & 25 & 10.1 & 3.408E-02 & 1.616E-03 & 4.707E-13 & 8.183E-04  \\
4 & 32 & 0.75  & -61 & 15 & 3 & 4.671E-03 & 6.217E-04 & 6.451E-14 & 1.122E-04  \\
4 & 30 & 17.05 & -61 & 18 & 44.1 & 5.260E-03 & 6.532E-04 & 7.264E-14 & 1.263E-04  \\
4 & 29 & 51.03 & -61 & 11 & 27.3 & 1.884E-03 & 4.234E-04 & 2.602E-14 & 4.524E-05  \\
4 & 30 & 41.88 & -61 & 17 & 9.9 & 6.554E-03 & 7.243E-04 & 9.051E-14 & 1.574E-05  \\
4 & 30 & 44.89 & -61 & 11 & 4.6 & 1.223E-03 & 3.569E-04 & 1.689E-14 & 2.937E-05  \\
4 & 30 & 41.86 & -61 & 13 & 3 & 2.163E-03 & 4.432E-04 & 2.987E-14 & 5.194E-05  \\
4 & 29 & 37.98 & -61 & 14 & 45.5 & 1.614E-03 & 3.951E-04 & 2.229E-14 & 3.875E-05 \\
4 & 29 & 32.16 & -61 & 13 & 13 & 1.587E-03 & 3.952E-04 & 2.192E-14 & 3.811E-05  \\
4 & 29 & 26.28 & -61 & 11 & 55.9 & 1.260E-03 & 3.650E-04 & 1.740E-14 & 3.025E-05  \\
4 & 29 & 2.02  & -61 & 13 & 54.9 & 1.559E-03 & 3.953E-04 & 2.153E-14 & 3.743E-05  \\
4 & 30 & 7.6   & -61 & 20 & 3.6 & 4.963E-03 & 6.353E-04 & 6.854E-14 & 1.192E-04  \\
4 & 30 & 33.1  & -61 & 24 & 22.6 & 1.586E-02 & 1.108E-03 & 2.190E-13 & 3.808E-04  \\
4 & 29 & 7.2   & -61 & 20 & 18.3 & 1.708E-03 & 4.023E-04 & 2.359E-14 & 4.101E-05 \\
4 & 29 & 12.95 & -61 & 15 & 8.4 & 1.058E-03 & 3.404E-04 & 1.461E-14 & 2.540E-05  \\
4 & 30 & 59.55 & -61 & 27 & 40.1 & 2.718E-02 & 1.444E-03 & 3.754E-13 & 6.526E-04 \\
4 & 29 & 51.96 & -61 & 18 & 12.9 & 4.704E-03 & 6.216E-04 & 6.496E-14 & 1.130E-04  \\
4 & 29 & 27.89 & -61 & 23 & 37 & 2.897E-03 & 4.981E-04 & 4.001E-14 & 6.956E-05 \\
4 & 30 & 3.09  & -61 & 17 & 14 & 4.020E-03 & 5.784E-04 & 5.552E-14 & 9.653E-05  \\
4 & 30 & 30.43 & -61 & 20 & 23.1 & 1.029E-02 & 8.983E-04 & 1.421E-13 & 2.471E-04  \\
4 & 30 & 41.87 & -61 & 27 & 15.8 & 1.648E-02 & 1.128E-03 & 2.276E-13 & 3.957E-04  \\
4 & 29 & 17.68 & -61 & 28 & 18.2 & 2.213E-03 & 4.430E-04 & 3.056E-14 & 5.314E-05  \\
4 & 30 & 33.23 & -61 & 33 & 0.7 & 4.172E-03 & 5.833E-04 & 5.762E-14 & 1.002E-04 \\
4 & 29 & 35.69 & -61 & 31 & 40.7 & 2.309E-03 & 4.494E-04 & 3.189E-14 & 5.544E-05  \\
4 & 30 & 21.56 & -61 & 28 & 48.7 & 9.564E-03 & 8.656E-04 & 1.321E-13 & 2.296E-04  \\
4 & 28 & 47.65 & -61 & 30 & 52.9 & 1.025E-03 & 3.316E-04 & 1.416E-14 & 2.461E-05  \\
4 & 30 & 53.73 & -61 & 30 & 5.7 & 1.048E-02 & 9.046E-04 & 1.447E-13 & 2.516E-04 \\
4 & 30 & 26.74 & -61 & 30 & 31.9 & 8.654E-03 & 8.247E-04 & 1.195E-13 & 2.078E-04  \\
4 & 28 & 56.57 & -61 & 37 & 53.8 & 1.535E-03 & 3.878E-04 & 2.120E-14 & 3.686E-05  \\
4 & 31 & 4.61  & -61 & 28 & 2.6 & 3.796E-02 & 1.704E-03 & 5.242E-13 & 9.115E-04  \\
4 & 30 & 28.84 & -61 & 39 & 32.5 & 1.995E-03 & 4.231E-04 & 2.755E-14 & 4.790E-05  \\
4 & 30 & 26.85 & -61 & 35 & 3.1 & 3.020E-03 & 5.038E-04 & 4.171E-14 & 7.252E-05  \\
4 & 31 & 12.64 & -61 & 27 & 28.9 & 5.944E-02 & 2.129E-03 & 8.209E-13 & 1.427E-03  \\
4 & 31 & 8.34  & -61 & 34 & 13.8 & 4.474E-03 & 6.027E-04 & 6.179E-14 & 1.074E-04  \\
4 & 31 & 7.7   & -61 & 30 & 12.1 & 1.268E-02 & 9.927E-04 & 1.751E-13 & 3.045E-04  \\
4 & 31 & 30.61 & -61 & 38 & 45 & 1.547E-03 & 3.800E-04 & 2.136E-14 & 3.715E-05  \\
4 & 30 & 50.84 & -61 & 35 & 11.7 & 2.267E-03 & 4.429E-04 & 3.131E-14 & 5.443E-05  \\
4 & 30 & 6.99  & -61 & 35 & 51.2 & 1.560E-03 & 3.800E-04 & 2.154E-14 & 3.746E-05  \\
4 & 31 & 39.3  & -61 & 39 & 51.6 & 1.606E-03 & 3.876E-04 & 2.218E-14 & 3.856E-05  \\
4 & 31 & 21.54 & -61 & 30 & 16.3 & 1.458E-02 & 1.063E-03 & 2.014E-13 & 3.501E-04  \\
4 & 31 & 17.87 & -61 & 35 & 37.9 & 3.477E-03 & 5.370E-04 & 4.802E-14 & 8.349E-05 \\
4 & 31 & 42.88 & -61 & 30 & 20.1 & 1.320E-02 & 1.013E-03 & 1.823E-13 & 3.170E-04  \\
4 & 32 & 29.88 & -61 & 30 & 44.6 & 5.414E-03 & 6.620E-04 & 7.477E-14 & 1.300E-04  \\
4 & 31 & 53.46 & -61 & 29 & 41.2 & 1.145E-02 & 9.451E-04 & 1.581E-13 & 2.749E-04 \\
4 & 32 & 4.87  & -61 & 28 & 49.9 & 1.569E-02 & 1.103E-03 & 2.168E-13 & 3.770E-04  \\
4 & 32 & 23.53 & -61 & 33 & 40 & 2.76E-003 & 4.86E-004 & 3.812E-14 & 6.627E-05  \\
4 & 32 & 58.46 & -61 & 30 & 5.1 & 2.488E-03 & 4.684E-04 & 3.436E-14 & 5.974E-05  \\
4 & 33 & 22.76 & -61 & 28 & 45.8 & 1.156E-03 & 3.487E-04 & 1.597E-14 & 2.776E-05  \\
4 & 31 & 53.72 & -61 & 26 & 14.4 & 2.640E-02 & 1.424E-03 & 3.646E-13 & 6.339E-04 \\
4 & 31 & 53.58 & -61 & 27 & 38.5 & 3.900E-02 & 1.728E-03 & 5.386E-13 & 9.365E-04  \\
4 & 32 & 57.83 & -61 & 25 & 47.2 & 4.231E-03 & 5.932E-04 & 5.843E-14 & 1.016E-04  \\
4 & 33 & 8.58  & -61 & 21 & 15.4 & 4.040E-03 & 5.835E-04 & 5.579E-14 & 9.701E-05  \\
4 & 32 & 54.13 & -61 & 21 & 32.4 & 5.583E-03 & 6.750E-04 & 7.710E-14 & 1.341E-04 \\
4 & 31 & 53.18 & -61 & 18 & 50 & 1.466E-02 & 1.068E-03 & 2.025E-13 & 3.520E-04  \\
4 & 31 & 48.33 & -61 & 16 & 33.8 & 9.482E-03 & 8.657E-04 & 1.310E-13 & 2.277E-04  \\
4 & 30 & 56.9  & -61 & 21 & 47.4 & 1.312E-02 & 1.010E-03 & 1.812E-13 & 3.150E-04  \\
4 & 31 & 9.75  & -61 & 25 & 35.8 & 3.568E-02 & 1.653E-03 & 4.928E-13 & 8.567E-04 \\
4 & 30 & 32.53 & -61 & 12 & 57.5 & 1.931E-03 & 4.232E-04 & 2.667E-14 & 4.637E-05  \\
4 & 30 & 39.66 & -61 & 17 & 51.9 & 7.007E-03 & 7.478E-04 & 9.677E-14 & 1.682E-04  \\
4 & 29 & 5.22  & -61 & 13 & 21.7 & 1.253E-03 & 3.650E-04 & 1.730E-14 & 3.009E-05 \\
4 & 29 & 18.93 & -61 & 14 & 19.3 & 1.207E-03 & 3.569E-04 & 1.667E-14 & 2.898E-05  \\
4 & 29 & 36.13 & -61 & 14 & 50.2 & 1.385E-03 & 3.726E-04 & 1.913E-14 & 3.326E-05 \\
4 & 30 & 16.68 & -61 & 22 & 39.4 & 9.999E-03 & 8.853E-04 & 1.381E-13 & 2.401E-04  \\
4 & 29 & 16.95 & -61 & 19 & 53.1 & 1.035E-03 & 3.316E-04 & 1.429E-14 & 2.485E-05  \\
4 & 30 & 12.3  & -61 & 19 & 46.4 & 5.343E-03 & 6.576E-04 & 7.379E-14 & 1.283E-04  \\
4 & 30 & 8.45  & -61 & 25 & 29.8 & 1.099E-02 & 9.267E-04 & 1.518E-13 & 2.639E-04 \\
4 & 31 & 13.24 & -61 & 27 & 12.2 & 6.460E-02 & 2.219E-03 & 8.922E-13 & 1.551E-03  \\
4 & 30 & 0.74  & -61 & 25 & 54.3 & 1.129E-02 & 9.390E-04 & 1.559E-13 & 2.711E-04 \\
4 & 30 & 25.99 & -61 & 27 & 45.9 & 1.177E-02 & 9.572E-04 & 1.626E-13 & 2.826E-04  \\
4 & 30 & 45.64 & -61 & 35 & 38.7 & 2.339E-03 & 4.494E-04 & 3.230E-14 & 5.616E-05  \\
4 & 30 & 58.25 & -61 & 35 & 36.2 & 2.341E-03 & 4.494E-04 & 3.233E-14 & 5.621E-05  \\
4 & 31 & 51.9  & -61 & 32 & 15.1 & 6.816E-03 & 7.361E-04 & 9.413E-14 & 1.637E-04  \\
4 & 31 & 17.66 & -61 & 27 & 43.7 & 7.082E-02 & 2.323E-03 & 9.781E-13 & 1.701E-03  \\
4 & 32 & 52.93 & -61 & 29 & 36.9 & 3.712E-03 & 5.582E-04 & 5.126E-14 & 8.913E-05  \\
4 & 32 & 25.62 & -61 & 23 & 20.0 & 1.353E-02 & 1.027E-03 & 1.869E-13 & 3.249E-04    \\
4 & 32 & 48.45 & -61 & 19 & 47.8 & 4.819E-03 & 6.309E-04 & 6.655E-14 & 1.157E-04  \\
4 & 31 & 26.94 & -61 & 23 & 1.3 & 3.544E-02 & 1.647E-03 & 4.894E-13 & 8.510E-04  \\
4 & 32 & 18.84 & -61 & 19 & 20.3 & 9.711E-03 & 8.756E-04 & 1.341E-13 & 2.332E-04  \\
4 & 31 & 48.16 & -61 & 18 & 2.1 & 1.208E-02 & 9.722E-04 & 1.668E-13 & 2.901E-04  \\
4 & 32 & 2.49  & -61 & 19 & 46.6 & 1.504E-02 & 1.082E-03 & 2.077E-13 & 3.611E-04  \\
4 & 32 & 21.93 & -61 & 10 & 41.2 & 1.176E-03 & 3.570E-04 & 1.624E-14 & 2.824E-05  \\
4 & 29 & 33.01 & -61 & 11 & 24.4 & 1.184E-03 & 3.570E-04 & 1.635E-14 & 2.843E-05  \\
4 & 30 & 46.27 & -61 & 11 & 32.2 & 1.306E-03 & 3.648E-04 & 1.804E-14 & 3.136E-05  \\
4 & 30 & 52.13 & -61 & 10 & 34.7 & 1.899E-03 & 4.233E-04 & 2.623E-14 & 4.560E-05  \\
4 & 30 & 18.94 & -61 & 18 & 6.9 & 5.482E-03 & 6.663E-04 & 7.571E-14 & 1.316E-04  \\
4 & 30 & 3.77  & -61 & 15 & 55.4 & 3.170E-03 & 5.208E-04 & 4.378E-14 & 7.612E-05  \\
4 & 30 & 4.38  & -61 & 22 & 55.5 & 6.880E-03 & 7.400E-04 & 9.502E-14 & 1.652E-04  \\
4 & 30 & 55.49 & -61 & 20 & 14.3 & 1.303E-02 & 1.007E-03 & 1.800E-13 & 3.129E-04 \\
4 & 30 & 38.92 & -61 & 28 & 37.5 & 1.298E-02 & 1.004E-03 & 1.793E-13 & 3.117E-04  \\
4 & 29 & 44.77 & -61 & 30 & 54.9 & 2.702E-03 & 4.804E-04 & 3.732E-14 & 6.488E-05  \\
4 & 31 & 0.91  & -61 & 29 & 57.9 & 1.101E-02 & 9.266E-04 & 1.521E-13 & 2.644E-04  \\
4 & 30 & 14.72 & -61 & 31 & 45.4 & 1.077E-02 & 9.173E-04 & 1.487E-13 & 2.586E-04 \\
4 & 30 & 32.07 & -61 & 42 & 0.6 & 1.511E-03 & 3.801E-04 & 2.087E-14 & 3.628E-05  \\
4 & 30 & 56.84 & -61 & 32 & 2.3 & 6.911E-03 & 7.400E-04 & 9.543E-14 & 1.659E-04 \\
4 & 31 & 4.72  & -61 & 30 & 53.4 & 8.431E-03 & 8.141E-04 & 1.164E-13 & 2.024E-04 \\ 
4 & 31 & 50.25 & -61 & 30 & 46.3 & 9.934E-03 & 8.821E-04 & 1.372E-13 & 2.385E-04 \\
\enddata
\end{deluxetable}

\startlongtable
\begin{deluxetable}{cccccccccccc}
    \tabletypesize{\scriptsize}
	\tablecaption{Calculated Values for Each Galaxy}
	\tablecolumns{14}
	\tablenum{4}
	\tablewidth{0pt}
	\tablehead{
		\colhead{} &
		\colhead{RA} &
		\colhead{} & \colhead{} & \colhead{Dec.} & \colhead{} & \colhead{R from cluster} & \colhead{Ion Density} & \colhead{Velocity} & \colhead{V$_{gal}$-V$_{cl}$} & \colhead{$\rho$} & \colhead{$\rho$ v$^{2}$} \\
		\colhead{hr} & \colhead{min} & \colhead{sec} & \colhead{deg} & \colhead{arcmin} & \colhead{arcsec} & \colhead{center (kpc)} & \colhead{(cm$^{-3}$)} & \colhead{(km s$^{-1}$)} & \colhead{(km/s)} & \colhead{(g/cm$^{3}$)} & \colhead{(dyn/cm$^{2}$)} 
	}
	\startdata
	4 & 31 & 50.03 & -61 & 19 & 29.7 & 572.91 & 2.7476E-03 & 15074 & -2730 & 5.9359E-27 & 4.4240E-10  \\
	4 & 30 & 34.31 & -61 & 19 & 12.6 & 700.735 & 2.0730E-03 & 16872 & -932 & 4.4784E-27 & 3.8901E-11  \\
	4 & 30 & 50.16 & -61 & 15 & 43.9 & 856.381 & 1.2916E-03 & 17054 & -750 & 2.7904E-27 & 1.5696E-11  \\
	4 & 31 & 40.34 & -61 & 16 & 35.3 & 753.041 & 1.8544E-03 & 15604 & -2200 & 4.0062E-27 & 1.9390E-10  \\
	4 & 30 & 59.59 & -61 & 25 & 7.2 & 242.725 & 3.8631E-03 & 20962 & 3158 & 8.3458E-27 & 8.3232E-10  \\
	4 & 30 & 8.03 & -61 & 18 & 9.3 & 915.99 & 1.4310E-03 & 20096 & 2292 & 3.0916E-27 & 1.6241E-10 \\
	4 & 31 & 39.39 & -61 & 21 & 41.8 & 387.482 & 3.7982E-03 & 20321 & 2517 & 8.2055E-27 & 5.1985E-10 \\
	4 & 31 & 13.2 & -61 & 10 & 37.9 & 1181.63 & 6.2505E-04 & 17929 & 125 & 1.3504E-27 & 2.1099E-13 \\
	4 & 32 & 47.12 & -61 & 21 & 7.2 & 836.301 & 1.5283E-03 & 20289 & 2485 & 3.3017E-27 & 2.0389E-10  \\
	4 & 33 & 14.12 & -61 & 20 & 34 & 1067.13 & 6.4343E-04 & 18316 & 512 & 1.3901E-27 & 3.6440E-12  \\
	4 & 32 & 55.41 & -61 & 19 & 10.4 & 974.164 & 1.0574E-03 & 16273 & -1531 & 2.2844E-27 & 5.3545E-11  \\
	4 & 32 & 26.65 & -61 & 21 & 34.2 & 665.097 & 2.0346E-03 & 17959 & 155 & 4.3956E-27 & 1.0560E-12 \\
	4 & 30 & 28.36 & -61 & 17 & 4 & 858.674 & 1.3857E-03 & 17547 & -257 & 2.9937E-27 & 1.9773E-12  \\
	4 & 29 & 44.51 & -61 & 14 & 56.3 & 1230 & 8.1135E-04 & 17565 & -239 & 1.7528E-27 & 1.0012E-12  \\
	4 & 29 & 57.45 & -61 & 11 & 45.7 & 1335.45 & 7.2191E-04 & 17551 & -253 & 1.5596E-27 & 9.9830E-13  \\
	4 & 30 & 56.87 & -61 & 13 & 33.5 & 991.857 & 8.0513E-04 & 16733 & -1071 & 1.7394E-27 & 1.9952E-11  \\
	4 & 30 & 58.75 & -61 & 10 & 21 & 1219.17 & 6.1985E-04 & 17178 & -626 & 1.3391E-27 & 5.2477E-12 \\
	4 & 30 & 51.24 & -61 & 10 & 59.7 & 1186.98 & 7.4303E-04 & 18044 & 240 & 1.6052E-27 & 9.2462E-13 \\
	4 & 29 & 39.59 & -61 & 10 & 34.5 & 1499.66 & 4.8643E-04 & 19795 & 1991 & 1.0509E-27 & 4.1658E-11 \\
	4 & 30 & 50.35 & -61 & 14 & 6 & 969.324 & 9.8085E-04 & 17101 & -703 & 2.1190E-27 & 1.0473E-11  \\
	4 & 29 & 5.27 & -61 & 13 & 2.4 & 1581.99 & 5.9283E-04 & 18756 & 952 & 1.2807E-27 & 1.1607E-11  \\
	4 & 30 & 4.47 & -61 & 23 & 25.7 & 739.598 & 1.9213E-03 & 17177 & -627 & 4.1508E-27 & 1.6318E-11  \\
	4 & 30 & 45.38 & -61 & 23 & 35.7 & 407.641 & 3.0177E-03 & 18839 & 1035 & 6.5193E-27 & 6.9837E-11 \\
	4 & 29 & 18.62 & -61 & 19 & 29.8 & 1224.15 & 7.0553E-04 & 17425 & -379 & 1.5242E-27 & 2.1894E-12  \\
	4 & 29 & 47.62 & -61 & 23 & 37 & 877.801 & 1.3562E-03 & 16984 & -820 & 2.9299E-27 & 1.9701E-11  \\
	4 & 30 & 22.67 & -61 & 23 & 19.6 & 593.014 & 2.8257E-03 & 18999 & 1195 & 6.1047E-27 & 8.7177E-11 \\
	4 & 29 & 24.75 & -61 & 19 & 5.7 & 1189.19 & 6.7687E-04 & 17202 & -602 & 1.4623E-27 & 5.2995E-12 \\
	4 & 30 & 53.08 & -61 & 25 & 14.2 & 291.649 & 3.6043E-03 & 15674 & -2130 & 7.7867E-27 & 3.5328E-10  \\
	4 & 30 & 28.68 & -61 & 19 & 39.8 & 708.259 & 2.0670E-03 & 16113 & -1691 & 4.4654E-27 & 1.2769E-10  \\
	4 & 29 & 32.77 & -61 & 14 & 52.5 & 1309.1 & 6.1871E-04 & 17159 & -645 & 1.3367E-27 & 5.5609E-12  \\
	4 & 29 & 27.18 & -61 & 16 & 51.9 & 1256.94 & 7.4367E-04 & 17112 & -692 & 1.6066E-27 & 7.6935E-12  \\
	4 & 31 & 8.71 & -61 & 27 & 7 & 139.94 & 5.9017E-03 & 15025 & -2779 & 1.2750E-26 & 9.8467E-10  \\
	4 & 29 & 2.33 & -61 & 24 & 17.2 & 1259.59 & 7.7298E-04 & 15782 & -2022 & 1.6700E-27 & 6.8276E-11 \\
	4 & 29 & 34.78 & -61 & 28 & 3.7 & 966.872 & 1.2418E-03 & 18397 & 593 & 2.6828E-27 & 9.4342E-12  \\
	4 & 29 & 33.1 & -61 & 15 & 36.3 & 1271.83 & 8.1767E-04 & 17601 & -203 & 1.7665E-27 & 7.2796E-13  \\
	4 & 30 & 20.03 & -61 & 26 & 9.7 & 564.658 & 2.4597E-03 & 15961 & -1843 & 5.3138E-27 & 1.8049E-10  \\
	4 & 29 & 24.92 & -61 & 26 & 3.4 & 1049.25 & 1.0054E-03 & 17096 & -708 & 2.1721E-27 & 1.0888E-11 \\
	4 & 29 & 25.36 & -61 & 33 & 45.9 & 1167.06 & 7.0670E-04 & 17578 & -226 & 1.5268E-27 & 7.7980E-13  \\
	4 & 30 & 37.69 & -61 & 29 & 2.7 & 444.872 & 2.5933E-03 & 18504 & 700 & 5.6025E-27 & 2.7452E-11  \\
	4 & 30 & 14.66 & -61 & 28 & 8.7 & 620.558 & 2.3676E-03 & 16478 & -1326 & 5.1150E-27 & 8.9936E-11  \\
	4 & 30 & 57.00 & -61 & 31 & 7.3 & 407.182 & 2.2869E-03 & 16743 & -1061 & 4.9407E-27 & 5.5618E-11 \\
	4 & 29 & 48.73 & -61 & 37 & 32.2 & 1159.08 & 7.0955E-04 & 18534 & 730 & 1.5329E-27 & 8.1689E-12  \\
	4 & 30 & 52.25 & -61 & 28 & 13.8 & 303.671 & 3.5168E-03 & 15484 & -2320 & 7.5976E-27 & 4.0893E-10  \\
	4 & 29 & 8.67 & -61 & 39 & 6.4 & 1500.71 & 7.2863E-04 & 18136 & 332 & 1.5741E-27 & 1.7351E-12  \\
	4 & 30 & 32.85 & -61 & 37 & 55.5 & 944.776 & 9.2432E-04 & 19680 & 1876 & 1.9969E-27 & 7.0278E-11  \\
	4 & 30 & 45.32 & -61 & 33 & 0.4 & 579.626 & 1.5958E-03 & 15332 & -2472 & 3.4476E-27 & 2.1068E-10  \\
	4 & 30 & 39.53 & -61 & 35 & 7.8 & 737.58 & 1.1947E-03 & 15664 & -2140 & 2.5810E-27 & 1.1820E-10  \\
	4 & 30 & 18.11 & -61 & 35 & 8.8 & 853.185 & 1.2882E-03 & 18188 & 384 & 2.7831E-27 & 4.1038E-12  \\
	4 & 30 & 37.57 & -61 & 42 & 49.6 & 1259.53 & 6.9207E-04 & 16720 & -1084 & 1.4952E-27 & 1.7569E-11  \\
	4 & 30 & 59.77 & -61 & 34 & 38.5 & 626.889 & 1.4766E-03 & 19966 & 2162 & 3.1901E-27 & 1.4911E-10 \\
	4 & 31 & 31.08 & -61 & 42 & 8.8 & 1143.39 & 4.3514E-04 & 16044 & -1760 & 9.4007E-28 & 2.9120E-11  \\
	4 & 31 & 17.25 & -61 & 31 & 49.8 & 387.17 & 2.3560E-03 & 15790 & -2014 & 5.0899E-27 & 2.0646E-10  \\
	4 & 31 & 42.26 & -61 & 31 & 39.2 & 402.426 & 2.1599E-03 & 16066 & -1738 & 4.6663E-27 & 1.4095E-10  \\
	4 & 32 & 13.9 & -61 & 38 & 12 & 956.743 & 6.8673E-04 & 18262 & 458 & 1.4836E-27 & 3.1121E-12  \\
	4 & 32 & 22.39 & -61 & 35 & 51 & 849.518 & 9.0024E-04 & 17605 & -199 & 1.9449E-27 & 7.7019E-13  \\
	4 & 32 & 33.84 & -61 & 38 & 38.8 & 1074.97 & 8.1098E-04 & 19422 & 1618 & 1.7521E-27 & 4.5867E-11 \\
	4 & 32 & 35.35 & -61 & 29 & 45.4 & 667.008 & 1.5294E-03 & 19592 & 1788 & 3.3041E-27 & 1.0563E-10  \\
	4 & 32 & 40.1 & -61 & 25 & 42.5 & 672.069 & 2.0146E-03 & 16942 & -862 & 4.3524E-27 & 3.2340E-11  \\
	4 & 31 & 55.17 & -61 & 30 & 42 & 405.088 & 2.2672E-03 & 17380 & -424 & 4.8981E-27 & 8.8056E-12  \\
	4 & 32 & 45.88 & -61 & 32 & 56.3 & 854.96 & 1.0279E-03 & 17385 & -419 & 2.2207E-27 & 3.8987E-12  \\
	4 & 32 & 42.82 & -61 & 28 & 11.1 & 701.442 & 1.5363E-03 & 18314 & 510 & 3.3190E-27 & 8.6326E-12  \\
	4 & 31 & 23.3 & -61 & 27 & 26.9 & 60.3938 & 6.3767E-03 & 17841 & 37 & 1.3776E-26 & 1.8860E-13  \\
	4 & 32 & 47.6 & -61 & 25 & 11.3 & 742.353 & 1.5193E-03 & 17305 & -499 & 3.2823E-27 & 8.1728E-12  \\
	4 & 32 & 13.28 & -61 & 25 & 40.2 & 438.438 & 3.1762E-03 & 16389 & -1415 & 6.8619E-27 & 1.3739E-10  \\
	4 & 32 & 41.58 & -61 & 23 & 3.8 & 730.969 & 1.8607E-03 & 17301 & -503 & 4.0199E-27 & 1.0171E-11 \\
	4 & 33 & 9.01 & -61 & 23 & 12.1 & 957.477 & 1.0573E-03 & 16301 & -1503 & 2.2841E-27 & 5.1598E-11  \\
	4 & 31 & 37.44 & -61 & 25 & 52.9 & 129.74 & 5.2210E-03 & 15324 & -2480 & 1.1279E-26 & 6.9373E-10  \\
	4 & 33 & 16.32 & -61 & 18 & 0.9 & 1175.16 & 4.8635E-04 & 17084 & -720 & 1.0507E-27 & 5.4469E-12  \\
	4 & 31 & 14.59 & -61 & 23 & 44.7 & 228.465 & 4.5582E-03 & 19346 & 1542 & 9.8475E-27 & 2.3415E-10  \\
	4 & 31 & 21.86 & -61 & 25 & 17.7 & 100.452 & 5.9436E-03 & 18944 & 1140 & 1.2841E-26 & 1.6688E-10  \\
	4 & 33 & 6.66 & -61 & 26 & 13.7 & 902.641 & 9.0863E-04 & 17843 & 39 & 1.9630E-27 & 2.9857E-14  \\
	4 & 32 & 54.18 & -61 & 25 & 47.7 & 794.861 & 1.3549E-03 & 15211 & -2593 & 2.9272E-27 & 1.9681E-10 \\
	4 & 33 & 15.31 & -61 & 21 & 43 & 1044.12 & 8.8472E-04 & 18418 & 614 & 1.9114E-27 & 7.2057E-12  \\
	4 & 31 & 58.7 & -61 & 16 & 24.1 & 812.513 & 1.6052E-03 & 18574 & 770 & 3.4679E-27 & 2.0561E-11  \\
	4 & 32 & 34.77 & -61 & 14 & 55.5 & 1063.63 & 8.5224E-04 & 18999 & 1195 & 1.8412E-27 & 2.6293E-11  \\
	4 & 32 & 34.08 & -61 & 13 & 59.1 & 1117.12 & 7.2260E-04 & 17160 & -644 & 1.5611E-27 & 6.4744E-12  \\
	4 & 30 & 51.87 & -61 & 23 & 24.4 & 369.989 & 3.0246E-03 & 17179 & -625 & 6.5344E-27 & 2.5525E-11  \\
	4 & 32 & 44.94 & -61 & 16 & 27 & 1034.41 & 8.4873E-04 & 20443 & 2639 & 1.8336E-27 & 1.2770E-10  \\
	4 & 30 & 41.79 & -61 & 22 & 43.1 & 470.951 & 2.8328E-03 & 16815 & -989 & 6.1199E-27 & 5.9860E-11  \\
	4 & 32 & 12.83 & -61 & 12 & 32.9 & 1122.36 & 7.7045E-04 & 17884 & 80 & 1.6645E-27 & 1.0653E-13  \\
	4 & 30 & 43.25 & -61 & 21 & 6.9 & 542.605 & 2.6529E-03 & 18739 & 935 & 5.7314E-27 & 5.0105E-11  \\
	4 & 30 & 54.21 & -61 & 21 & 3.2 & 487.828 & 2.8990E-03 & 18554 & 750 & 6.2629E-27 & 3.5229E-11 \\
	4 & 31 & 8.81 & -61 & 25 & 10.1 & 172.409 & 4.6737E-03 & 19241 & 1437 & 1.0097E-26 & 2.0850E-10  \\
	4 & 32 & 0.75 & -61 & 15 & 3 & 911.775 & 1.2637E-03 & 17017 & -787 & 2.7300E-27 & 1.6909E-11  \\
	4 & 30 & 17.05 & -61 & 18 & 44.1 & 828.953 & 1.5530E-03 & 16940 & -864 & 3.3552E-27 & 2.5046E-11  \\
	4 & 29 & 51.03 & -61 & 11 & 27.3 & 1386.88 & 7.0946E-04 & 17729 & -75 & 1.5327E-27 & 8.6216E-14  \\
	4 & 30 & 41.88 & -61 & 17 & 9.9 & 790.061 & 1.7026E-03 & 16800 & -1004 & 3.6783E-27 & 3.7078E-11 \\
	4 & 30 & 44.89 & -61 & 11 & 4.6 & 1196.17 & 5.9927E-04 & 17619 & -185 & 1.2947E-27 & 4.4310E-13 \\
	4 & 30 & 41.86 & -61 & 13 & 3 & 1066.98 & 8.5190E-04 & 17094 & -710 & 1.8405E-27 & 9.2777E-12  \\
	4 & 29 & 37.98 & -61 & 14 & 45.5 & 1280.83 & 7.0591E-04 & 18316 & 512 & 1.5251E-27 & 3.9979E-12  \\
	4 & 29 & 32.16 & -61 & 13 & 13 & 1396.9 & 6.6070E-04 & 16861 & -943 & 1.4274E-27 & 1.2693E-11  \\
	4 & 29 & 26.28 & -61 & 11 & 55.9 & 1501.01 & 5.5977E-04 & 19447 & 1643 & 1.2093E-27 & 3.2645E-11  \\
	4 & 29 & 2.02 & -61 & 13 & 54.9 & 1564.73 & 6.1738E-04 & 17650 & -154 & 1.3338E-27 & 3.1632E-13  \\
	4 & 30 & 7.6 & -61 & 20 & 3.6 & 829.675 & 1.5355E-03 & 16615 & -1189 & 3.3172E-27 & 4.6897E-11  \\
	4 & 30 & 33.1 & -61 & 24 & 22.6 & 478.578 & 3.1260E-03 & 16253 & -1551 & 6.7535E-27 & 1.6246E-10 \\
	4 & 29 & 7.2 & -61 & 20 & 18.3 & 1292.92 & 7.5746E-04 & 18086 & 282 & 1.6364E-27 & 1.3014E-12 \\
	4 & 29 & 12.95 & -61 & 15 & 8.4 & 1433.28 & 5.4089E-04 & 19138 & 1334 & 1.1685E-27 & 2.0795E-11  \\
	4 & 30 & 59.55 & -61 & 27 & 40.1 & 228.912 & 4.2247E-03 & 16020 & -1784 & 9.1271E-27 & 2.9049E-10  \\
	4 & 29 & 51.96 & -61 & 18 & 12.9 & 1021.56 & 1.3736E-03 & 16924 & -880 & 2.9674E-27 & 2.2980E-11  \\
	4 & 29 & 27.89 & -61 & 23 & 37 & 1046.75 & 1.1219E-03 & 19230 & 1426 & 2.4237E-27 & 4.9286E-11  \\	
	4 & 30 & 3.09 & -61 & 17 & 14 & 994.406 & 1.2662E-03 & 19803 & 1999 & 2.7354E-27 & 1.0931E-10  \\
	4 & 30 & 30.43 & -61 & 20 & 23.1 & 659.584 & 2.3187E-03 & 17422 & -382 & 5.0093E-27 & 7.3098E-12  \\
	4 & 30 & 41.87 & -61 & 27 & 15.8 & 374.249 & 3.2775E-03 & 18851 & 1047 & 7.0806E-27 & 7.7618E-11  \\
	4 & 29 & 17.68 & -61 & 28 & 18.2 & 1118.19 & 9.7861E-04 & 16611 & -1193 & 2.1142E-27 & 3.0090E-11  \\
	4 & 30 & 33.23 & -61 & 33 & 0.7 & 647.975 & 1.6065E-03 & 19579 & 1775 & 3.4706E-27 & 1.0935E-10  \\
	4 & 29 & 35.69 & -61 & 31 & 40.7 & 1022.05 & 1.0567E-03 & 19537 & 1733 & 2.2830E-27 & 6.8565E-11  \\
	4 & 30 & 21.56 & -61 & 28 & 48.7 & 572.642 & 2.4576E-03 & 19556 & 1752 & 5.3094E-27 & 1.6297E-10  \\
	4 & 28 & 47.65 & -61 & 30 & 52.9 & 1409.58 & 5.8696E-04 & 16577 & -1227 & 1.2681E-27 & 1.9091E-11  \\
	4 & 30 & 53.73 & -61 & 30 & 5.7 & 368.955 & 2.6218E-03 & 19566 & 1762 & 5.6642E-27 & 1.7585E-10 \\
	4 & 30 & 26.74 & -61 & 30 & 31.9 & 579.922 & 2.3443E-03 & 17322 & -482 & 5.0646E-27 & 1.1766E-11  \\
	4 & 28 & 56.57 & -61 & 37 & 53.8 & 1536.47 & 6.8442E-04 & 16438 & -1366 & 1.4786E-27 & 2.7591E-11  \\
	4 & 31 & 4.61 & -61 & 28 & 2.6 & 200.33 & 4.9934E-03 & 18436 & 632 & 1.0788E-26 & 4.3089E-11  \\
	4 & 30 & 28.84 & -61 & 39 & 32.5 & 1066.38 & 9.5169E-04 & 18443 & 639 & 2.0560E-27 & 8.3952E-12  \\
	4 & 30 & 26.85 & -61 & 35 & 3.1 & 797.672 & 1.3116E-03 & 18954 & 1150 & 2.8335E-27 & 3.7473E-11 \\
	4 & 31 & 12.64 & -61 & 27 & 28.9 & 118.546 & 6.2351E-03 & 17815 & 11 & 1.3470E-26 & 1.6299E-14  \\
	4 & 31 & 8.34 & -61 & 34 & 13.8 & 575.962 & 1.6543E-03 & 16828 & -976 & 3.5739E-27 & 3.4044E-11  \\
	4 & 31 & 7.7 & -61 & 30 & 12.1 & 299.563 & 2.8826E-03 & 19017 & 1213 & 6.2275E-27 & 9.1629E-11  \\
	4 & 31 & 30.61 & -61 & 38 & 45 & 893.592 & 8.5656E-04 & 19018 & 1214 & 1.8505E-27 & 2.7273E-11  \\
	4 & 30 & 50.84 & -61 & 35 & 11.7 & 694.436 & 1.1577E-03 & 19051 & 1247 & 2.5011E-27 & 3.8893E-11  \\
	4 & 30 & 6.99 & -61 & 35 & 51.2 & 958.211 & 8.9319E-04 & 17831 & 27 & 1.9297E-27 & 1.4067E-14  \\
	4 & 31 & 39.3 & -61 & 39 & 51.6 & 982.534 & 8.3394E-04 & 17263 & -541 & 1.8016E-27 & 5.2731E-12  \\
	4 & 31 & 21.54 & -61 & 30 & 16.3 & 268.719 & 3.0735E-03 & 16782 & -1022 & 6.6400E-27 & 6.9353E-11  \\
	4 & 31 & 17.87 & -61 & 35 & 37.9 & 664.515 & 1.4117E-03 & 20052 & 2248 & 3.0498E-27 & 1.5412E-10  \\
	4 & 31 & 42.88 & -61 & 30 & 20.1 & 318.514 & 2.8475E-03 & 18340 & 536 & 6.1518E-27 & 1.7674E-11  \\
	4 & 32 & 29.88 & -61 & 30 & 44.6 & 652.299 & 1.5851E-03 & 16247 & -1557 & 3.4244E-27 & 8.3016E-11 \\
	4 & 31 & 53.46 & -61 & 29 & 41.2 & 342.161 & 2.5989E-03 & 18611 & 807 & 5.6146E-27 & 3.6565E-11  \\
	4 & 32 & 4.87 & -61 & 28 & 49.9 & 393.206 & 2.9550E-03 & 18754 & 950 & 6.3841E-27 & 5.7616E-11  \\
	4 & 32 & 23.53 & -61 & 33 & 40 & 734.949 & 1.1251E-03 & 17724 & -80 & 2.4306E-27 & 1.5556E-13  \\
	4 & 32 & 58.46 & -61 & 30 & 5.1 & 867.206 & 9.6310E-04 & 17327 & -477 & 2.0807E-27 & 4.7341E-12  \\
	4 & 33 & 22.76 & -61 & 28 & 45.8 & 1054.78 & 5.9493E-04 & 18867 & 1063 & 1.2853E-27 & 1.4523E-11  \\
	4 & 31 & 53.72 & -61 & 26 & 14.4 & 262.149 & 3.9098E-03 & 17975 & 171 & 8.4467E-27 & 2.4699E-12   \\
	4 & 31 & 53.58 & -61 & 27 & 38.5 & 269.631 & 4.7922E-03 & 17485 & -319 & 1.0353E-26 & 1.0535E-11   \\
	4 & 32 & 57.83 & -61 & 25 & 47.2 & 826.926 & 1.2479E-03 & 17649 & -155 & 2.6960E-27 & 6.4772E-13 \\
	4 & 33 & 8.58 & -61 & 21 & 15.4 & 1001.73 & 1.1099E-03 & 17968 & 164 & 2.3978E-27 & 6.4492E-13  \\
	4 & 32 & 54.13 & -61 & 21 & 32.4 & 877.067 & 1.3830E-03 & 16675 & -1129 & 2.9877E-27 & 3.8083E-11  \\
	4 & 31 & 53.18 & -61 & 18 & 50 & 628.685 & 2.5394E-03 & 17516 & -288 & 5.4861E-27 & 4.5504E-12  \\
	4 & 31 & 48.33 & -61 & 16 & 33.8 & 771.326 & 1.9301E-03 & 20996 & 3192 & 4.1698E-27 & 4.2485E-10  \\
	4 & 30 & 56.9 & -61 & 21 & 47.4 & 429.472 & 2.7580E-03 & 17310 & -494 & 5.9584E-27 & 1.4541E-11  \\
	4 & 31 & 9.75 & -61 & 25 & 35.8 & 147.512 & 4.7968E-03 & 18530 & 726 & 1.0363E-26 & 5.4620E-11  \\
	4 & 30 & 32.53 & -61 & 12 & 57.5 & 1104.62 & 7.9732E-04 & 15292 & -2512 & 1.7225E-27 & 1.0870E-10  \\
	4 & 30 & 39.66 & -61 & 17 & 51.9 & 754.95 & 1.7965E-03 & 17315 & -489 & 3.8811E-27 & 9.2805E-12  \\
	4 & 29 & 5.22 & -61 & 13 & 21.7 & 1567.39 & 5.5071E-04 & 16185 & -1619 & 1.1898E-27 & 3.1185E-11 \\
	4 & 29 & 18.93 & -61 & 14 & 19.3 & 1428.65 & 5.7492E-04 & 16192 & -1612 & 1.2421E-27 & 3.2275E-11  \\
	4 & 29 & 36.13 & -61 & 14 & 50.2 & 1288.91 & 6.5256E-04 & 17003 & -801 & 1.4098E-27 & 9.0453E-12  \\
	4 & 30 & 16.68 & -61 & 22 & 39.4 & 661.843 & 2.3545E-03 & 18279 & 475 & 5.0866E-27 & 1.1477E-11  \\
	4 & 29 & 16.95 & -61 & 19 & 53.1 & 1225.5 & 6.0471E-04 & 18302 & 498 & 1.3064E-27 & 3.2400E-12  \\
	4 & 30 & 12.3 & -61 & 19 & 46.4 & 809.427 & 1.5966E-03 & 18852 & 1048 & 3.4492E-27 & 3.7883E-11  \\
	4 & 30 & 8.45 & -61 & 25 & 29.8 & 670.808 & 2.5304E-03 & 16346 & -1458 & 5.4666E-27 & 1.1621E-10  \\
	4 & 31 & 13.24 & -61 & 27 & 12.2 & 104.311 & 6.4950E-03 & 18343 & 539 & 1.4032E-26 & 4.0765E-11  \\
	4 & 30 & 0.74 & -61 & 25 & 54.3 & 735.28 & 2.5267E-03 & 17973 & 169 & 5.4586E-27 & 1.5590E-12 \\
	4 & 30 & 25.99 & -61 & 27 & 45.9 & 517.688 & 2.7389E-03 & 17443 & -361 & 5.9171E-27 & 7.7112E-12  \\	
	4 & 30 & 45.64 & -61 & 35 & 38.7 & 744.156 & 1.1621E-03 & 18143 & 339 & 2.5105E-27 & 2.8851E-12  \\
	4 & 30 & 58.25 & -61 & 35 & 36.2 & 698.054 & 1.1683E-03 & 18741 & 937 & 2.5239E-27 & 2.2159E-11  \\
	4 & 31 & 51.9 & -61 & 32 & 15.1 & 480.197 & 1.9777E-03 & 18967 & 1163 & 4.2726E-27 & 5.7789E-11  \\
	4 & 31 & 17.66 & -61 & 27 & 43.7 & 98.4949 & 6.7977E-03 & 18931 & 1127 & 1.4686E-26 & 1.8653E-10  \\
	4 & 32 & 52.93 & -61 & 29 & 36.9 & 810.925 & 1.2046E-03 & 17893 & 89 & 2.6024E-27 & 2.0613E-13  \\
	4 & 32 & 25.62 & -61 & 23 & 20 & 593.535 & 2.4529E-03 & 17121 & -683 & 5.2992E-27 & 2.4720E-11 \\
	4 & 32 & 48.45 & -61 & 19 & 47.8 & 897.545 & 1.2703E-03 & 18819 & 1015 & 2.7444E-27 & 2.8273E-11  \\
	4 & 31 & 26.94 & -61 & 23 & 1.3 & 266.986 & 4.5657E-03 & 15589 & -2215 & 9.8638E-27 & 4.8394E-10  \\
	4 & 32 & 18.84 & -61 & 19 & 20.3 & 721.781 & 1.9621E-03 & 17349 & -455 & 4.2390E-27 & 8.7757E-12 \\
	4 & 31 & 48.16 & -61 & 18 & 2.1 & 667.423 & 2.2770E-03 & 17878 & 74 & 4.9193E-27 & 2.6938E-13  \\
	4 & 32 & 2.49 & -61 & 19 & 46.6 & 607.527 & 2.5797E-03 & 15928 & -1876 & 5.5732E-27 & 1.9614E-10  \\
	4 & 32 & 21.93 & -61 & 10 & 41.2 & 1279.99 & 5.3243E-04 & 19706 & 1902 & 1.1503E+27 & 4.1612E-11  \\
	4 & 29 & 33.01 & -61 & 11 & 24.4 & 1489.4 & 5.4236E-04 & 16971 & -833 & 1.1717E-27 & 8.1304E-12  \\
	4 & 30 & 46.27 & -61 & 11 & 32.2 & 1160.23 & 6.2950E-04 & 16057 & -1747 & 1.3600E-27 & 4.1506E-11  \\
	4 & 30 & 52.13 & -61 & 10 & 34.7 & 1214.9 & 7.3551E-04 & 17785 & -19 & 1.5890E-27 & 5.7363E-15  \\
	4 & 30 & 18.94 & -61 & 18 & 6.9 & 850.29 & 1.5619E-03 & 19080 & 1276 & 3.3743E-27 & 5.4939E-11  \\
	4 & 30 & 3.77 & -61 & 15 & 55.4 & 1059.89 & 1.0825E-03 & 18261 & 457 & 2.3387E-27 & 4.8843E-12  \\
	4 & 30 & 4.38 & -61 & 22 & 55.5 & 753.058 & 1.9110E-03 & 17431 & -373 & 4.1286E-27 & 5.7440E-12  \\
	4 & 30 & 55.49 & -61 & 20 & 14.3 & 533.91 & 2.6462E-03 & 16672 & -1132 & 5.7168E-27 & 7.3256E-11  \\
	4 & 30 & 38.92 & -61 & 28 & 37.5 & 423.35 & 2.9105E-03 & 20213 & 2409 & 6.2879E-27 & 3.6491E-10  \\
	4 & 29 & 44.77 & -61 & 30 & 54.9 & 927.796 & 1.1858E-03 & 16408 & -1396 & 2.5617E-27 & 4.9923E-11  \\
	4 & 31 & 0.91 & -61 & 29 & 57.9 & 318.831 & 2.6887E-03 & 20087 & 2283 & 5.8087E-27 & 3.0275E-10  \\
	4 & 30 & 14.72 & -61 & 31 & 45.4 & 716.612 & 2.5449E-03 & 17011 & -793 & 5.4981E-27 & 3.4574E-11  \\
	4 & 30 & 32.07 & -61 & 42 & 0.6 & 1220.01 & 7.6823E-04 & 16941 & -863 & 1.6597E-27 & 1.2361E-11  \\
	4 & 30 & 56.84 & -61 & 32 & 2.3 & 464.289 & 2.1128E-03 & 19685 & 1881 & 4.5644E-27 & 1.6150E-10  \\
	4 & 31 & 4.72 & -61 & 30 & 53.4 & 356.561 & 2.3469E-03 & 19538 & 1734 & 5.0701E-27 & 1.5245E-10 \\
	4 & 31 & 50.25 & -61 & 30 & 46.3 & 381.508 & 2.4270E-03 & 17140 & -664 & 5.2432E-27 & 2.3117E-11  \\
	\enddata:
\end{deluxetable}

\end{document}